\documentclass[graybox]{svmult}

\usepackage{mathptmx}       
\usepackage{helvet}         
\usepackage{courier}        
\usepackage{type1cm}        
\usepackage{makeidx}         
\usepackage{graphicx}        
\usepackage{multicol}        
\usepackage[bottom]{footmisc}
\usepackage{amsmath}
\usepackage{bm}

\def\x{{\bf{x}}}

\def\P{{\bf{P}}}

\def\f{{\bf{f}}}

\def\fm{\hat{{\bf{f}}}}
\def\gm{\hat{{\bf{g}}}}
\def\lam{{\bm{\lambda}}}
\def\lamm{\hat{{\bm{\lambda}}}}

\def\xm{\hat{{\bf{x}}}}
\def\ym{\hat{{\bf{y}}}}

\makeindex             

\begin{document}

\title*{Deterministic treatment of model error in geophysical data assimilation}
\author{Alberto Carrassi and St\'ephane Vannitsem}
\institute{Alberto Carrassi \at NERSC - Nansen Environmental and Remote Sensing Center, Bergen, Norway, \email{alberto.carrassi@nersc.no}
\and St\'ephane Vannitsem \at RMI - Royal Meteorological Institute of Belgium, Brussels, Belgium \email{svn@meteo.be}}
%

%
\maketitle

\abstract*{
This chapter describes a novel approach for the treatment of model error in geophysical data assimilation. In this method, model error is treated as a deterministic process fully correlated in time. This allows for the derivation of the evolution equations for the relevant moments of the model error statistics required in data assimilation procedures, along with an approximation suitable for application to large numerical models typical of environmental science. In this contribution we first derive the equations for the model error dynamics in the general case, and then for the particular situation of parametric error.
We show how this deterministic description of the model error can be incorporated in sequential and variational data assimilation procedures. A numerical comparison with standard methods is given using low-order dynamical systems, prototypes of atmospheric circulation, and a realistic soil model.
The deterministic approach proves to be very competitive with only minor additional computational cost. Most importantly, it offers a new way to address the problem of accounting for model error in data assimilation that can easily be implemented in systems of increasing complexity and in the context of modern ensemble-based procedures. }

\abstract{This chapter describes a novel approach for the treatment of model error in geophysical data assimilation. In this method, model error is treated as a deterministic process fully correlated in time. This allows for the derivation of the evolution equations for the relevant moments of the model error statistics required in data assimilation procedures, along with an approximation suitable for application to large numerical models typical of environmental science. In this contribution we first derive the equations for the model error dynamics in the general case, and then for the particular situation of parametric error.
We show how this deterministic description of the model error can be incorporated in sequential and variational data assimilation procedures. A numerical comparison with standard methods is given using low-order dynamical systems, prototypes of atmospheric circulation, and a realistic soil model.
The deterministic approach proves to be very competitive with only minor additional computational cost. Most importantly, it offers a new way to address the problem of accounting for model error in data assimilation that can easily be implemented in systems of increasing complexity and in the context of modern ensemble-based procedures. }

\section{Introduction}
\label{sec:1}

The prediction problem in geophysical fluid dynamics typically relies on two complementary elements: the model and the data. The mathematical model, and its discretized version, embodies our knowledge about the laws governing the system evolution, while the data are samples of the system's state. They give complementary information about the same object. The sequence of operations that merges model and data to obtain a possibly improved estimate of the flow’s state is usually known, in environmental science, as data assimilation \cite{Daley1991,Kalnay2002}. The physical and dynamical complexity of geophysical systems makes the data assimilation problem particularly involved.

The different information entering the data assimilation procedure, usually the model, the data and a background field representing the state estimate prior to the assimilation of new observations, are weighted according to their respective accuracy. Data assimilation in geophysics, particularly in numerical weather prediction (NWP) has experienced a long and fruitful stream of research in recent decades which has led to a number of advanced methods able to take full advantage of the increasing amount of available observations and to efficiently track and reduce the dynamical instabilities \cite{Evensen}. As a result the overall accuracy of the Earth’s system estimate and prediction, particularly the atmosphere, has improved dramatically.

Despite this trend of improvement, the treatment of model error in data assimilation procedures is still, in most instances, done following simple assumptions such as the absence of time correlation \cite{Jazwinski}. The lack of attention on model error is in part justified by the fact that on the time scale of NWP, where most of the geophysical data assimilation advancements have been originally concentrated, its influence is reasonably considered small as compared to the initial condition error that grows in view of the chaotic nature of the dynamics. Nevertheless, the improvement in data assimilation techniques and observational networks on the one hand, and the recent growth of interest in seasonal-to-decadal prediction on the other \cite{DobRey-et-al-13, Weber_et_al_2015}, has placed model error, and its treatment in data assimilation, as a main concern and a key priority. A number of studies reflecting this concern have appeared, in the context of sequential and variational schemes \cite{DeeDaSilva1998,Tremolet2006,Tremolet2007, Kondrashov2008}.

Two main obstacles toward the development of techniques taking into account model error sources are the huge size of the geophysical models and the wide range of possible model error sources. The former problem implies the need to estimate large error covariance matrices on the basis of the limited number of available observations. The second important issue is related to the multiple sources of modeling error, such as incorrect parametrisation, numerical discretization, and the lack of description of some relevant scale of motion. This latter problem has until recently limited the development of a general formulation for the model error dynamics. Model error is commonly $modeled$ as an additive, stationary, zero-centered, Gaussian white noise process. This choice could be legitimate by the multitude of unknown error sources and the central limit theorem. However, despite this simplification, the size of geoscientific models still makes detailed estimation of the stochastic model error covariance impractical. 

In the present contribution we describe an alternative approach in which the evolution of the model error is described based on a deterministic short-time approximation. The approximation is suitable for realistic applications and is used to estimate the model error contribution in the state estimate. The method is based on the theory of deterministic dynamics of the model error that was introduced recently by \cite{Nicolis2003,Nicolis2004,Nicolis2009}.  
Using this approach it is possible to derive evolution equations for the moments of the model error statistics required in data assimilation procedures, and has been applied in the context of both sequential and variational data assimilation schemes, and for errors originated from uncertain parameters and from unresolved scales. 

We give here a review of the recent developments of the deterministic treatment of model error in data assimilation. To this end, we start by first formalizing the deterministic model error dynamics in Sect. 2. We show how general equations for the mean and covariance error can be obtained and discuss the parametric error as a special case. In Sections 3 and 4 the incorporation of the short-time model error evolution laws is described in the context of the Extended Kalman filter and variational scheme respectively. These two types of assimilation procedures are significantly different and are summarized in the respective Sections along with the discussion on the consequences of the implementation of the model error treatment. We provide some numerical illustrations of the proposed approach together with comparisons with other methods, for two prototypical low order chaotic systems widely used in theoretical studies in geosciences \cite{Lorenz1963,Lorenz1996} and a quasi-operational soil model \cite{Mahfouf2007}. 

New potential applications of the use of the deterministic model error treatment are currently under way and are summarized, along with a synopsis of the method, in the final discussion Section 5. These include soil data assimilation with the use of new observations and ensemble based procedures \cite{Evensen}.

\section{Formulation\label{sec:formulation}}
\label{sec:2}

Let the model at our disposal be represented as:
\begin{equation}
\label{EQM}
\frac{d\x(t)}{dt} = \f(\x, \lam),
\end{equation}
where $\f$ is typically a nonlinear function, defined in ${\mathbf R}^N$ and $\lam$ is a $P$-dimensional vector of parameters.

Model (\ref{EQM}) is used to describe the evolution of a (unknown) true dynamics,  {\it i.e. nature}, whose evolution is assumed to be given by the following coupled equations:
\begin{equation} 
\label{nature}
 \frac{d\xm (t)}{dt} = \fm(\xm,\ym,\lamm) \qquad \frac{d\ym (t)}{dt} = \gm(\xm,\ym,\lamm) 
\end{equation} 
where $\xm$ is a vector in ${\mathbf R}^N$, and $\ym$ is defined in ${\mathbf R}^L$ and may represent scales that are present in the real world, but are neglected in model (\ref{EQM}); the unknown parameters $\lamm$ have dimension $P$. The true state is thus a vector of dimension $N+L$. The model state vector $\x$ and the variable $\xm$ of the true dynamics span the same phase space although, given the difference in the functions $\f$ and $\fm$, they do not have the same attractor in general. 
The function ${\bf f}$ can have an explicit dependence on time but it is dropped here to simplify the notation. 

When using model (\ref{EQM}) to describe the evolution of $\xm$, estimation error can arise from the uncertainty in the initial conditions at the resolved scale ($\x(t_0)\ne \xm(t_0)$) and from the approximate description of the nature afforded by (\ref{EQM}) which is referred as model error. A number of different sources of model errors are present in environmental modeling. Typical examples are those arising from the inadequate description of some physical processes, numerical discretization and/or the presence of scales in the actual dynamics that are unresolved by the model. The latter are typically parametrised in terms of the resolved variables (for instance the Reynolds stress of the turbulent transport).   

\subsection{General description of model error dynamics}

Following the approach outlined in \cite{Nicolis2004}, we derive the evolution equations of the dominant moments, mean and covariance, of the estimation error $\delta\x = \x - \xm$ in the resolved scale ({\it i.e.} in ${\mathbf R}^N$). The formal solutions of (\ref{EQM}) and (\ref{nature}) read respectively:
\begin{equation}  
\label{modsol}
\x(t) = \x_0 + \int_0^t d\tau \f(\x(\tau),\lam)
\end{equation}
\begin{equation}  
\label{trusol}
\xm(t) =\xm_0 + \int_0^t d\tau \fm(\xm(\tau),\ym(\tau),\lamm)
\end{equation}
where $\x_0=\x(t_0)$, and $\xm_0 = \xm(t_0)$.
By taking the difference between (\ref{modsol}) and (\ref{trusol}), and averaging over an ensemble of perturbations around a reference state, we get the formal solution for the mean error, the bias:
\begin{equation}
\label{bias}
<\delta\x(t)> = < \delta\x_0> + \int_0^t d\tau < \f(\x(\tau),\lam) -  \fm(\xm(\tau),\ym(\tau),\lamm) >
\end{equation}
with  $\delta\x_0 = \x_0 - \xm_0$. Two types of averaging could be performed, one over a set of initial conditions sampled on the attractor
of the system, and/or a set of perturbations around one specific initial state selected on the system's attractor. In data assimilation, the
second is more relevant since one is interested in the local evaluation of the uncertainty. However, in many situations the first one is used
to get statistical information on covariances quantities, as will be illustrated in this Chapter. For clarity, we will refer to $<.>$ as the
local averaging, and to $<<.>>$ for an averaging over a set of initial conditions sampled over the attractor of the system. In this section,
we will only use $<.>$ for clarity, but it also extends to the other averaging. We will use the other notation $<<.>>$ when necessary.  

In the hypothesis that the initial condition is unbiased, $<\delta\x_0>=0$, Eq. (\ref{bias}) gives the evolution equation of the bias due to the model error, usually refers to as drift in climate prediction context. The important factor driving the drift is the difference between the true and modeled tendency fields, $< \f(\x(\tau),\lam) -  \fm(\xm(\tau),\ym(\tau),\lamm) >$. Expanding (\ref{bias}) in Taylor series around $t_0=0$ up to the first non-trivial order, and using unbiased initial conditions, it reads:
\begin{equation}
\label{bias-TC}
{\bf b}^m(t)=<\delta\x(t)> \approx < \f(\x(\tau),\lam) -  \fm(\xm(\tau),\ym(\tau),\lamm) >t
\end{equation}
Equation (\ref{bias-TC}) gives the evolution of the bias, ${\bf b}^m$, the drift, in the short-time approximation and the subscript $m$ stands for model error-related bias. It is important to remark that in the case of stochastic model error treatment, and in the hypothesis of unbiased initial condition error, ${\bf b}^m = 0$. 
  
Similarly, by taking the expectation of the external product of the error anomalies $\delta\x$ by themselves, we have:
\begin{equation}
{\bf P}(t) = < \{ \delta\x(t)\}\{\delta\x(t) \}^T> = < \{ \delta\x_0\}\{\delta\x_0 \}^T> + \nonumber  
\end{equation}
\begin{equation}
<\{ \delta\x_0\} \{\int_0^t d\tau [ \f(\x(\tau),\lam) - \fm(\xm(\tau),\ym(\tau),\lamm) ] \}^T > +  \nonumber 
\end{equation}
\begin{equation}
<\{\int_0^t d\tau  [ \f(\x(\tau),\lam) - \fm(\xm(\tau),\ym(\tau),\lamm) ] \} \{ \delta\x_0\}^T > + \nonumber 
\end{equation}
\begin{equation}
\label{cov}
\int_0^t d\tau \int_0^t d\tau^{'} <\{ \f(\x(\tau),\lam) - \fm(\xm(\tau),\ym(\tau),\lamm)\} \{ \f (\x(\tau^{'}),\lam) - \fm(\xm(\tau^{'}),\ym(\tau^{'}),\lamm)\}^T >
\end{equation}

Equation (\ref{cov}) describes the time evolution of the estimation error covariance in the resolved scale. The first term, that does not depend on time, represents the covariance of the initial error. The two following terms account for the correlation between the error in the initial condition and the model error, while the last term combines the effect of both errors on the evolution of the estimation error covariance. 

Let us focus on the last term of Eq. (\ref{cov}) denoted as,
\begin{equation}
\label{cov-mod}
{\bf P}(t) = \int_0^t d\tau \int_0^t d\tau^{'} <\{ \f(\x(\tau),\lam) - \fm(\xm(\tau),\ym(\tau),\lamm)\} \{ \f(\x(\tau^{'}),\lam) - \fm(\xm(\tau^{'}),\ym(\tau^{'}),\lamm)\}^T >
\end{equation}
The amplitude and structure of this covariance depends on the dynamical properties of the difference of the nature and model tendency fields.    
Assuming that these differences are correlated in time, we can expand (\ref{cov-mod}) in a time series up to the first nontrivial order around the arbitrary initial time $t_0=0$, and gets:
\begin{equation}
\label{cov_aprx}
{\bf P}^m(t) \approx <\{ \f(\x_0,\lam) - \fm(\xm_0,\ym_0,\lamm)\} \{ \f(\x_0,\lam) - \fm(\xm_0,\ym_0,\lamm)\}^T > t^2 = {\bf Q}t^2 
\end{equation}
where ${\bf Q}$ is the model error covariance matrix at initial time. Note again that, if the terms ${\bf f}-\hat{{\bf f}}$ are represented as white-noise process, the short-time evolution of ${\bf P}(t)$ is bound to be linear instead of quadratic. 
This distinctive feature is relevant in data assimilation applications where model error is often assumed to be uncorrelated in time, a choice allowing for a reduction of the computational cost associated with certain types of algorithms \cite{Tremolet2006,CV10}. 

\subsection{Model error due to parameter uncertainties}

We assume for simplicity that the model resolves all scales present in the reference system.
Under the aforementioned hypothesis that the model and the true trajectories span the same phase space, nature dynamics, (\ref{nature}), can be rewritten as: 
\begin{equation}
\label{truth2}
\frac{d\xm (t)}{dt} = \f(\xm,\lamm)  + \epsilon {\bf h}(\xm,{\bf \gamma})
\end{equation}
The function ${\bf h}$, which has the same order of magnitude of $\f$ and is scaled by the dimensionless parameter $\epsilon$, accounts for all other extra terms not included in the model and depends on the resolved variable $\xm$ and on a set of additional parameters ${\bf \gamma}$. In a more formal description, this ${\bf h}$ would correspond to a function relating the variables $\xm$ and $\ym$ under an adiabatic elimination \cite{Nicolis2004}. We are interested here in a situation in which the main component of the nature dynamics is well captured by the model so that $\epsilon << 1$, and the extra terms described by ${\bf h}$ are neglected. We concentrate in a situation in which model error is due only to uncertainties in the specification of the parameters appearing in the evolution law $\f$. This formulation accounts, for instance, for errors in the description of some physical processes (dissipation, external forcing, etc.) represented by the parameters.

An equation for the evolution of the state estimation error $\delta{\bf x}$ can be obtained by taking the difference between the first rhs term in (\ref{truth2}) and (\ref{EQM}). The evolution of $\delta{\bf x}$ depends on the error estimate at the initial time $t=t_0$ (initial condition error $\delta{\bf x}(t_0)=\delta{\bf x}_0$) and on the model error. If $\delta{\bf x}$ is "small", the linearized dynamics provides a reliable approximation of the actual error evolution. The linearization is made along a model trajectory, solution of (\ref{EQM}), by expanding, to first order in $\delta{\bf x}$ and $\delta\lam =\lam-\lamm$, the difference between Eqs. (\ref{truth2}) and (\ref{EQM}):
\begin{equation}
\label{dyn-err-deriv}
\frac{d\delta{\bf x}}{dt} = \frac{\partial\f}{\partial\x}|_{\x}\delta{\bf x} + \frac{\partial\f}{\partial\lam}|_{\lam}\delta\lam   
\end{equation}
The first partial derivative on the rhs of (\ref{dyn-err-deriv}) is the Jacobian of the model dynamics evaluated along its trajectory. The second term, which corresponds to the model error, will be denoted $\delta\bm\mu$ hereafter to simplify the notation; $\delta\bm\mu= \frac{\partial\f}{\partial\lam}|_{\lam}\delta\lam$

The solution of (\ref{dyn-err-deriv}), with initial condition $\delta{\bf x}_0$ at $t=t_0$, reads:
\begin{equation}
\delta{\bf x}(t) = {\bf M}_{t,t_0}\delta{\bf x}_0 + \int^{t}_{t_{0}} d\tau{\bf M}_{t,\tau}\delta{\bf \mu}(\tau) \nonumber  
\end{equation}
\begin{equation}
\label{dyn-err-solution}
= \delta{\bf x}^{ic}(t) + \delta{\bf x}^{m}(t) 
\end{equation}

with ${\bf M}_{t,t_0}$ being the fundamental matrix (the propagator) relative to the linearized dynamics along the trajectory between $t_0$ and $t$. We point out that $\delta\bm\mu$ and ${\bf M}_{t,\tau}$ in (\ref{dyn-err-solution}) depend on $\tau$ (the integration variable) through the state variable ${\bf x}$. Equation (\ref{dyn-err-solution}) states that, in the linear approximation, the error in the state estimate is given by the sum of two terms, the evolution of initial condition error, $\delta{\bf x}^{ic}$, and the model error, $\delta{\bf x}^{m}$. 
The presence of the fundamental matrix ${\bf M}$ in the expression for $\delta{\bf x}^{m}$ suggests that the instabilities of the flow plays a role in the dynamics of model error. 

Let us now apply the expectation operator to (\ref{dyn-err-solution}) defined locally around the reference trajectory, by sampling over an ensemble of initial conditions and model errors, and the equation for the mean estimation error along a reference trajectory reads:    
\begin{equation}
<\delta{\bf x}(t)> = {\bf M}_{t,t_0}<\delta{\bf x}_0> + \int^{t}_{t_{0}} d\tau{\bf M}_{t,\tau}<\delta\bm\mu(\tau)> \nonumber   
\end{equation}
\begin{equation}
\label{dyn-err-solution-mean}
= <\delta{\bf x}^{ic}> + <\delta{\bf x}^m> 
\end{equation}

In a perfect model scenario an unbiased state estimate at time $t_0$ ($<\delta{\bf x}_0>=0$) will evolve, under the linearized dynamics, into an unbiased estimate at time $t$. In the presence of model error and, depending on its properties, an initially unbiased estimate can evolve into a biased one with $<\delta\bm\mu(t)>$ being the key factor.

The dynamics of the state estimation error covariance matrix can be obtained by taking the expectation of the outer product of $\delta{\bf x}(t)$ with itself. Assuming that the estimation error bias is known and removed from the background error, we get:

\begin{equation}
{\bf P}(t) = < \delta{\bf x}(t)\delta{\bf x}(t)^T > \nonumber  
\end{equation}
\begin{equation}
\label{dyn-cov-general}
=  {\bf P}^{ic}(t) + {\bf P}^{m}(t) + {\bf P}^{corr}(t) + ({\bf P}^{corr})^T(t)
\end{equation}
where:
\begin{equation}
\label{PIC}
{\bf P}^{ic}(t) = {\bf M}_{t,t_0} < \delta{\bf x}_0{\delta{\bf x}_0}^T > {\bf M}^T_{t,t_0} 
\end{equation}
\begin{equation}
\label{PMOD}
{\bf P}^{m}(t) =  \int^{t}_{t_{0}} d\tau \int^{t}_{t_{0}} d\tau^{'} {\bf M}_{t,\tau}< \delta\bm\mu(\tau){\delta\bm\mu(\tau^{'})}^T)> {\bf M}^T_{t,\tau^{'}}
\end{equation}
\begin{equation}
\label{PCORR}
{\bf P}^{corr}(t) = {\bf M}_{t,t_0}<(\delta{\bf x}_0)\left( \int^{t}_{t_{0}}d\tau{\bf M}_{t,\tau}\delta\bm\mu(\tau)\right)^T> 
\end{equation}
The four terms of the r.h.s. of (\ref{dyn-cov-general}) give the contribution to the estimation error covariance at time $t$  due to the initial condition, model error and their cross correlation, respectively. These integral equations are of little practical use for any realistic nonlinear systems, let alone the big models used in environmental prediction. A suitable expression can be obtained by considering their short-time approximations through a Taylor expansion around $t_0$. 
We proceed by expanding (\ref{dyn-err-solution}) in Taylor series, up to the first non trivial order, only for the model error term $\delta{\bf x}^m$ while keeping the initial condition term, $\delta{\bf x}^{ic}$, unchanged.  
In this case, the model error $\delta{\bf x}^{m}$ evolves linearly with time according to:
\begin{equation}
\label{dyn-err-mod_Ib}
\delta{\bf x}^{m} \approx \delta\bm\mu_0(t-t_0) 
\end{equation}
where $\delta\bm\mu(t_0)=\delta\bm\mu_0$.

By adding the initial condition error term, $\delta{\bf x}^{ic}$, we get a short time approximation of (\ref{dyn-err-solution}):
\begin{equation}
\label{dyn-err-der_Ib}
\delta{\bf x}(t) \approx {\bf M}_{t,t_0}\delta{\bf x}_0 + \delta\bm\mu_0(t-t_0) 
\end{equation}
For the mean error we get:
\begin{equation}
\label{errore-medio}
{\bf b}^m(t) \approx <\delta{\bf x}(t)> \approx {\bf M}_{t,t_0}<\delta{\bf x}_0> + <\delta\bm\mu_0>(t-t_0) 
\end{equation}
Therefore, as long as $<\delta\bm\mu_0>$ is different from zero, the bias due to parametric error evolves linearly for short-time, otherwise the evolution is conditioned by higher orders of the Taylor expansion. 
Note that the two terms in the short time error evolution (\ref{dyn-err-der_Ib}) and (\ref{errore-medio}), are not on equal footing since, in contrast to the model error term, which has been expanded up to the first nontrivial order in time, the initial condition error evolution contains all the orders of times $(t,t^2,...,t^n)$. The point is that, as explained below, we intend to use these equations to model the error evolution in conjunction with the technique of data assimilation for which the full matrix ${\bf M}$, or an amended ensemble based approximation, is already available.

Taking the expectation value of the external product of (\ref{dyn-err-der_Ib}) by itself and averaging, we get:
\begin{equation}
{\bf P}(t) \approx {\bf M}_{t,t_0} < \delta{\bf x}_0{\delta{\bf x}_0}^T > {\bf M}_{t,t_0}^T + \nonumber 
\end{equation}
\begin{equation}
\label{P-approx}
+ [ < \delta\bm\mu_0{\delta{\bf x}_0}^T >{\bf M}^T_{t,t_0} 
+ {\bf M}_{t,t_0} < \delta{\bf x}_0{\delta\bm\mu_0}^T > ](t-t_0)  
+  < \delta\bm\mu_0{\delta\bm\mu_0}^T > (t-t_0)^2
\end{equation}

Equation (\ref{P-approx}) is the short time evolution equation, in this linearized setting, for the error covariance matrix in the presence of both initial condition and parametric model errors.


\section{Deterministic model error treatment in the extended Kalman filter \label{sec:EKF}}

We describe here two formulations of the extended Kalman filter (EKF) incorporating a model error treatment. The Short-Time-Extended-Kalman-Filter, ST-EKF \cite{CVN08} accounts for model error through an estimate of its contribution to the assumed forecast error statistics. In the second formulation, the Short-Time-Augmented-Extended-Kalman-Filter, ST-AEKF \cite{CV11a}, the state estimation in the EKF is accompanied with the estimation of the uncertain parameters. This is done in the context of a general framework known as state augmentation \cite{Jazwinski}. In both cases model error is treated as a deterministic process implying that the dynamical laws described in the previous section are incorporated, at different stages, in the filter formulations. 

The EKF extends, to nonlinear dynamics, the classical Kalman filter (KF) for linear dynamics \cite{Kalman1960}. The algorithm is sequential in the sense that a prediction of the system's state is updated at discrete times, when observations are present. The state update, the analysis, is then taken as the initial condition for the subsequent prediction up to the next observation time. 
The EKF, as well as the standard KF for linear dynamics, is derived in the hypothesis of Gaussian errors whose distributions can thus be fully described using only the first two moments, the mean and the covariance. Although this can represent a very crude approximation, especially for nonlinear systems, it allows for a dramatic reduction of the cost and difficulties involved in the time propagation of the full error distribution. 

The model equations can conveniently be written in terms of a discrete mapping from time $t_k$ to $t_{k+1}$:
\begin{equation}
\label{model-dyn-discrete}
{\bf x}^{f}_{k+1} = {\mathcal M}{\bf x}^{a}_k
\end{equation}
where ${\bf x}^{f}$ and ${\bf x}^{a}$ are the forecast and analysis states respectively and ${\mathcal M}$ is the nonlinear model forward operator (the resolvent of (\ref{EQM})).

Let us assume that a set of $M$ noisy observations of the true system (\ref{nature}), stored as the components of an $M$-dimensional observation vector ${\bf y}^o$, is available at the regularly spaced discrete times $t_k=t_0+k\tau$, $k=1,2...$, with $\tau$ being the assimilation interval, so that:
\begin{equation}
\label{obs}
{\bf y}^o_k={\mathcal H}(\xm_k)+{\bf \epsilon^o_k}
\end{equation}
where ${\bf \epsilon^o}$ is the observation error, assumed here to be Gaussian with known covariance matrix ${\bf R}$ and uncorrelated in time. ${\mathcal H}$ is the (possibly nonlinear) observation operator which maps from model to observation space ({\it i.e.} from model to observed variables) and may involve spatial interpolations as well as transformations based on physical laws for indirect measurements \cite{JanicCohn06}.
 
For the EKF, as well as for most least-square based assimilation schemes, the analysis state update equation at an arbitrary analysis time $t_{k}$, reads \cite{Jazwinski}:
\begin{equation}
\label{analysis-update}
{\bf x}^{a}=\left[{\bf I} -{\bf K}{\bf H}\right]{\bf x}^{f} + {\bf K}{\bf y}^{o}  
\end{equation}
where the time indexes are dropped to simplify the notation. The analysis error covariance, ${\bf P}^{a}$, is updated through:
\begin{equation}
\label{analysis-update-P}
{\bf P}^{a}=\left[{\bf I} -{\bf K}{\bf H}\right]{\bf P}^{f}
\end{equation}
The $I\times M$ {\it gain} matrix ${\bf K}$ is given by:
\begin{equation}
\label{gain matrix}
{\bf K}={\bf P}^{f}{\bf H}^T\left[{\bf H}{\bf P}^{f}{\bf H}^T + {\bf R}\right]^{-1}  
\end{equation}
where ${\bf P}^{f}$ is the $I\times I$ forecast error covariance matrix and ${\bf H}$ the linearized observation operator (a $M\times I$ real matrix). 
The analysis update is thus based on two complementary sources of information, the observations, ${\bf y}^o$, and the forecast ${\bf x}^{f}$. The errors associated to each of them are assumed to be uncorrelated and fully described by the covariance matrices ${\bf R}$ and ${\bf P}^f$ respectively.   

In the EKF, the forecast error covariance matrix, ${\bf P}^f$, is obtained by linearizing the model around its trajectory between two successive analysis times $t_{k}$ and $t_{k+1}=t_{k}+\tau$. 
In the standard formulation of the EKF model error is assumed to be a random uncorrelated noise whose effect is modeled by adding a model error covariance matrix, ${\bf P}^m$, at the forecast step so that \cite{Jazwinski}:
\begin{equation}
\label{Pf_EKF}
{\bf P}^f = {\bf M}{\bf P}^a{\bf M}^T + {\bf P}^{m}
\end{equation}
In practice the matrix ${\bf P}^m$ should be considered as a measure of the variability of the noise sequence. This approach has been particularly attractive in the past in view of its simplicity and because of the lack of more refined model for the model error dynamics. Note that while ${\bf P}^f$ is propagated in time and is therefore flow dependent, ${\bf P}^m$ is defined once for all and it is then kept constant.

\subsection{Short Time Extended Kalman Filter - ST-EKF}

We study here the possibility of estimating the model error covariance, ${\bf P}^m$, on a deterministic basis \cite{CVN08}. The approach uses the formalism on model error dynamics outlined in Sect. 2. 

Model error is regarded as a time-correlated process and the short-time evolution laws (\ref{bias-TC}) and (\ref{cov_aprx}) are used to estimate the bias, ${\bf b}^m$, and the model error covariance matrix, ${\bf P}^m$, respectively. 
The adoption of the short-time approximation is also legitimated by the sequential nature of the EKF, and an important practical concern is the ratio between the duration of the short-time regime and the length of the assimilation interval $\tau$ over which the approximation is used \cite{Nicolis2004}. 

A key issue is the estimation of the two first statistical moments of the tendency mismatch, $\f -\fm$, required in (\ref{bias-TC}) and in (\ref{cov_aprx}) respectively. 
The problem is addressed assuming that a reanalysis dataset of relevant geophysical fields is available and is used as a proxy of the nature evolution. Reanalysis programs constitute the best-possible estimate of the Earth system over an extended period of time, using an homogeneous model and data assimilation procedure, and are of paramount importance in climate diagnosis (see {\it e.g.} \cite{Dee2011}). 

Let us suppose to have access to such a reanalysis which includes the analysis, ${\bf x}^a_r$, and the forecast field, ${\bf x}^f_r$, so that ${\bf x}^f_r(t_j+\tau_r) = \mathcal{M}{\bf x}^a_r (t_{j})$, and $\tau_r$ is the assimilation interval of the data assimilation scheme used to produce the reanalysis; the suffix $r$ stands for reanalysis. 
Under this assumption the following approximation is made:
\begin{equation}
{\bf f}({\bf x},\lam) - \hat{{\bf f}}(\hat{{\bf x}},\hat{{\bf y}},\lam,\hat{{\bf \epsilon}}) = \frac{d{\bf x}}{dt} - \frac{d\hat{{\bf x}}}{dt} \approx  \nonumber
\end{equation}
\begin{equation}
\label{deriv-apx}   
\frac{{\bf x}^f_r(t+\tau_r) - {\bf x}^a_r(t) }{\tau_r} - \frac{{\bf x}^a_r(t+\tau_r) - {\bf x}^a_r(t) }{\tau_r} = \frac{{\bf x}^f_r(t+\tau_r) - {\bf x}^a_r(t+\tau_r) }{\tau_r} = - \frac{\delta{\bf x}^a_r}{\tau_r}
\end{equation}
The difference between the analysis and the forecast, $\delta{\bf x}^a_r$, is usually referred, in data assimilation literature, to as the {\it analysis increment}. From (\ref{deriv-apx}) we see that the vector of analysis increments can be used to estimate the difference between the model and the true tendencies. A similar approach was originally introduced by Leith (1978) \cite{Leith1978}, and it has been used recently to account for model error in data assimilation \cite{Li2009}. 

Note that the estimate (\ref{deriv-apx}) neglects the analysis error, so that its accuracy is connected to that of the data assimilation algorithm used to produce the reanalysis, which is in turn related to the characteristics of the observational network such as number, distribution and frequency of the observations. 
However this error is present and acts as an initial condition error, a contribution which is already accounted for in the EKF update by the forecast error covariance, $\P^f$. As a consequence when (\ref{cov_aprx}) is used to estimate only the model error component, an overestimation is expected that can be overcome by an optimal tuning of the amplitude of ${\bf b}^m$ and ${\bf P}^m$.

The most straightforward way to estimate the bias due to model error using (\ref{deriv-apx}) in (\ref{bias-TC}), so that at analysis time it reads:
\begin{equation}
\label{bias-TC-2}
{\bf b}^m = - \sqrt{\alpha}<\delta\x^a_r>\frac{\tau}{\tau_r}
\end{equation}
The bias is then removed from the forecast field before the latter enters the EKF analysis update, (\ref{analysis-update}). The scalar term $\alpha$ is a tunable coefficient aimed at optimizing the bias size to account for the expected overestimation connected with the use of (\ref{deriv-apx}).
In a similar way the model error contribution to the forecast error covariance can be estimated taking the external product of (\ref{deriv-apx}) after removing the mean and reads:
\begin{equation}
\label{cov_aprx_3}
{\bf P}^m = \alpha<\{\delta{\bf x}^a_r - <\delta{\bf x}^a_r> \} \{\delta{\bf x}^a_r - <\delta{\bf x}^a_r>    \}^T > \frac{\tau^2}{\tau_r^2} \end{equation}

We consider now the particular case of parametric error. The forecast error covariance ${\bf P}^f$, is estimated using the short-time evolution (\ref{P-approx}) where the correlation terms are neglected and the model error covariance, ${\bf P}^m$ is evolved quadratically in the intervals between observations. An additional advantage is that ${\bf P}^m$ can be straightforwardly adapted to different assimilation intervals and for the assimilation of asynchronous observations. 
At analysis times the forecast error bias due to the model error, ${\bf b}^m$, can be estimated on the basis of the short-time approximation (\ref{errore-medio}):
\begin{equation}
\label{bias-PM}
{\bf b}^m= <\delta{\bf x}^m> \approx <\delta\bm\mu_o>\tau
\end{equation} 
By neglecting the correlation terms and dropping the time dependence for convenience, Eq. (\ref{P-approx}) can be rewritten as:
\begin{equation}
\label{PF-DETER}
{\bf P}^f = {\bf MP}^a{\bf M}^T + <\delta\bm\mu_o\delta\bm\mu_o^T>\tau^2 = {\bf MP}^a{\bf M}^T + {\bf Q}\tau^2= {\bf MP}^a{\bf M}^T + {\bf P}^m
\end{equation}
where ${\bf P}^a$ is the analysis error covariance matrix, as estimated at the last analysis time, and 
\begin{equation}
\label{PPM}
{\bf P}^m = {\bf Q}\tau^2 = <\delta\bm\mu_o\delta\bm\mu_o^T>\tau^2. 
\end{equation}

An essential ingredient of the ST-EKF in the case of parametric error is the matrix ${\bf Q}$: it embeds the information on the model error through the unknown parametric error $\delta\lam$ and the parametric functional dependence of the dynamics. 
In \cite{CVN08} it was supposed that some a-priori information on the model error was at disposal and could be used to prescribe $<\delta\bm\mu_o>$ and ${\bf Q}$ then used to compute ${\bf b}^m$ and ${\bf P}^m$ required by the ST-EKF. The availability of information on the model error, which may come in practice from the experience of modelers, is simulated by estimating $<\delta\bm\mu_o>$ and ${\bf Q}$ averaging over a large sample of states on the system's attractor as,
\begin{eqnarray}
{\bf b}^m & = & <<\delta\bm\mu_o>> \\
{\bf Q} & = & <<\delta\bm\mu_o\delta\bm\mu_o^T>> 
\end{eqnarray}
The same assumption is adopted here in the numerical applications with the ST-EKF described in Sect. 3.2.1.

In summary, in the ST-EKF, either in general or in the parameteric error case, once ${\bf b}^m$ and ${\bf P}^m$ are estimated (with (\ref{bias-TC-2})-(\ref{cov_aprx_3}) or (\ref{bias-PM})-(\ref{PPM}) respectively) they are then kept constant along the entire assimilation cycle.
Model error is thus repeatedly corrected in the subspace spanned by the range of ${\bf P}^m$ where it is supposed to be confined.
This choice reflects the assumption that the impact of model uncertainty on the forecast error does not fluctuate too much along the analysis cycle. 
Finally, in the ST-EKF, the forecast field and error covariance are transformed according to:
\begin{equation}
\label{ST-EKF-b}
\x^f \Longrightarrow \x^f - {\bf b}^m,
\end{equation}
\begin{equation}
\label{ST-EKF-P}
\P^f \Longrightarrow \P^f + {\bf P}^m
\end{equation}
These new first guess and forecast error covariance, (\ref{ST-EKF-b}) and (\ref{ST-EKF-P}), are then used in the EKF analysis formulas (\ref{analysis-update})-(\ref{analysis-update-P}).

\subsubsection{Numerical Results with ST-EKF. Error due to unresolved scale}

We show here numerical results of the ST-EKF for the case of model error arising from the lack of description of a scale of motion. The case of parametric error is numerically tested in Sect. 3.2.1.
A standard approach, known in geosciences as observation system simulation experiments (OSSE), is adopted here \cite{Bengt81}. This experimental setup is based on a twin model configuration in which a trajectory, solution of the system taken to represent the actual dynamics, is sampled to produce synthetic observations. A second model provides the trajectory that assimilates the observations.

As a prototype of two-scales chaotic dynamics we consider the model introduced by \cite{Lorenz1996}, whose equations read:
\begin{equation}
\label{model-expr_I}
\frac{dx_i}{dt}=(x_{i+1}-x_{i-2})x_{i-1} -  x_i + F - \frac{hc}{b}\sum_{j=1}^{10}y_{j,i}, \qquad  i=\{1,...,36\}
\end{equation}
\begin{equation}
\label{model-expr_II}
\frac{dy_{j,i}}{dt}=-cby_{j+1,i}(y_{j+2,i}-y_{j-1,i}) -  cy_{j,i} +  \frac{hc}{b}x_{i}, \qquad  j=\{1,...,10\}
\end{equation}
The model possesses two distinct scales of motion evolving according to (\ref{model-expr_I}) and (\ref{model-expr_II}), respectively. The large/slow scale variable, $x_i$, represents a generic meteorological variable over a circle at fixed latitude. In both set of equations, the quadratic term simulates the advection, the second rhs term the internal dissipation, while the constant term in (\ref{model-expr_I}) plays the role of the external forcing. The two scales are coupled in such a way that the small/fast scale variables $y_{j,i}$ inhibit the larger ones, while the opposite occurs for the effect of the variables $x_i$ on $y_{j,i}$. According to \cite{Lorenz1996} the variables $y_{j,i}$ can be taken to represent some convective-scale quantity, while the variables $x_i$ favor this convective activity. The model parameters are set as in \cite{Lorenz1996}: $c=b=10$, which makes the variables $x_{i}$ to vary ten times slower than $y_{j,i}$, with amplitudes ten times larger, while $F=10$ and $h=1$. With this choice, the dynamics is chaotic. The numerical integration have been performed using a fourth-order Runge-Kutta scheme with a time step of 0.0083 units, corresponding to 1 hour of simulated time.

In the experiments the full equations (\ref{model-expr_I}) - (\ref{model-expr_II}), are taken to represent the truth, while the model sees only the slow scale and its equations are given by (\ref{model-expr_I}) without the last term. A network of $M=12$ regularly spaced noisy observations of ${\bf x}$ is simulated by sampling the reference true trajectory  and adding a Gaussian random observation error.
We first generate a long record of analysis for the state vector, ${\bf x}$, which constitutes the reanalysis dataset. The EKF algorithm is run for $10$ years with assimilation interval $\tau_r=6$ hours, and observation variance set to $5\%$ of the system's climate variance. From this long integration we extract the record of analysis increments required in (\ref{bias-TC-2}) and (\ref{cov_aprx_3}). 

\begin{figure}[h]
\begin{center}
\includegraphics[height=6.4cm,width=9.5cm]{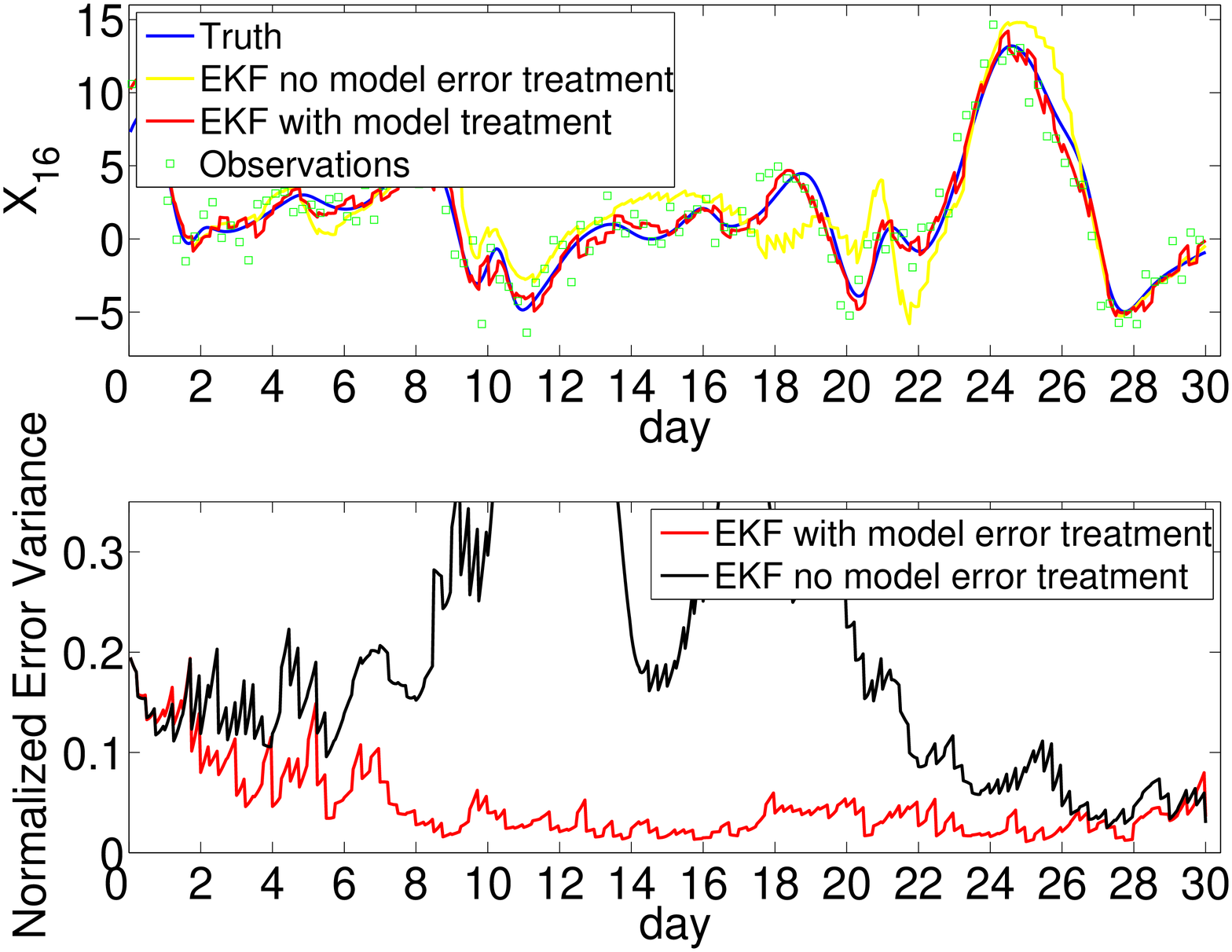}
\end{center}
\caption{Top Panel: Model variable $\hat{x}_{16}$ for the truth (blue), EKF without model error treatment (yellow), EKF with model error treatment (red) and observations (green). Bottom Panel: Estimation error variance, normalized with respect to the system's climate variance, as a function of time. From \cite{CV11b}.} 
\label{fig1}
\end{figure}

An illustration of the impact of the proposed treatment of the model error is given in Fig. \ref{fig1}, which shows a $30$ days long assimilation cycle. The upper panel displays the true large scale variable $x_{16}$ (blue line), the corresponding estimates obtained with the ST-EKF and the EKF without the model error treatment (red and yellow lines respectively) and the observations (green marks). The error variance of the EKF estimates are shown in the bottom panel.
From the top panel we see the improvement in the tracking of the true trajectory obtained by implementing the proposed model error treatment; this is particularly evident in the proximity of the maxima and minima of the true signal. The benefit is further evident by looking at the estimated error variance which undergoes a rapid convergence to values close or below the observation error.

\begin{figure}[h]
\begin{center}
\includegraphics[height=5cm,width=10cm]{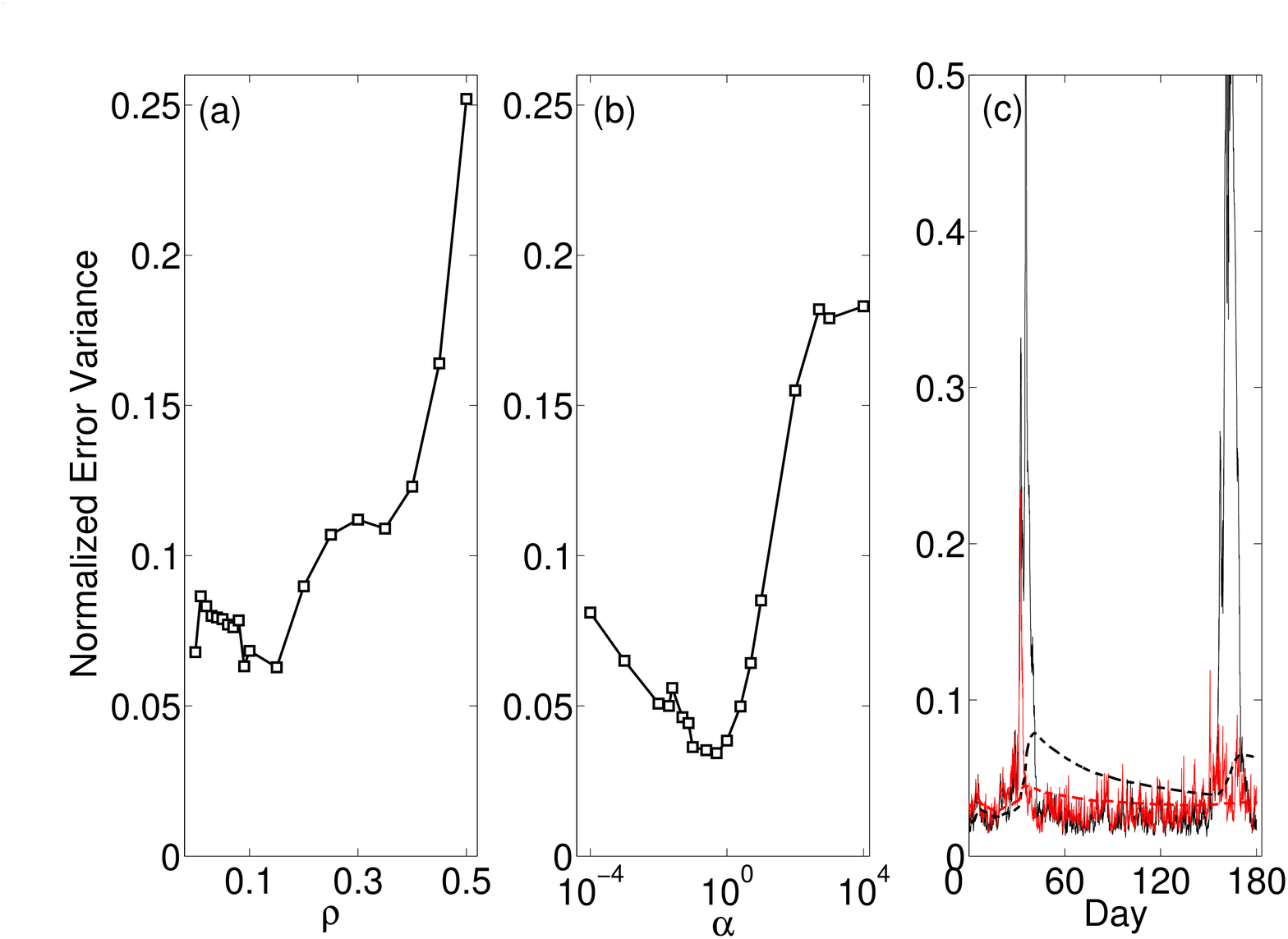}
\end{center}
\caption{Averaged normalized estimation error variance as a function of (a) the inflation factor $\rho$, (b) the coefficient $\alpha$ (log scale in the x-axis), and (c) time evolution of the normalized estimation error variance for the case $\rho=0.09$ (black) and $\alpha=0.5$ (red) (the time running mean is displayed with dashed lines). From \cite{CV11b}. }
\label{fig2}
\end{figure}

A common practical procedure used to account for model error in KF-like and ensemble-based schemes, is the multiplicative covariance inflation \cite{Anderson1999}. The forecast error covariance matrix ${\bf P}^f$ is multiplied by a scalar factor and thus inflated while keeping its spatial structure unchanged, so that ${\bf P}^f \rightarrow (1+\rho){\bf P}^f$ before its use in the analysis update, (\ref{analysis-update}). We have followed the same procedure here and have optimized the EKF by tuning the inflation factor $\rho$; the results are reported in Fig. \ref{fig2}(a) which shows the normalized estimation error variance as a function of $\rho$. 
The experiments last for $210$ days, and the results are averaged in time, after an initial transient of $30$ days, and over a sample of $100$ random initial conditions. The best performance is obtained by inflating ${\bf P}^f$ by $9\%$ of its original amplitude and the estimation error variance is about $6\%$ of the system's climate variance, slightly above the observation error variance. Note that when $\rho=0$ filter divergence occurs in some of the $100$ experiments.

We now test the sensitivity of the ST-EKF to the multiplicative coefficient $\alpha$ in (\ref{bias-TC-2}) and (\ref{cov_aprx_3}). The results are reported in Fig. \ref{fig2}(b), which shows the estimation error variance as a function of $\alpha$. As above the averages are taken over $180$ days and over the same ensemble of $100$ random initial conditions.  
The important feature is the existence of a range of values of $\alpha$, for which the estimation error is below the observation error level.
Note that for $\alpha=1$, the estimation error is about $4\%$ of the climate's variance, below the observational accuracy. 
This result highlights the accuracy of the estimate of ${\bf P}^m$ despite the simplifying assumptions such as the one associated with the correlation between model error and initial condition error and the use of the reanalysis field as a proxy of the actual true trajectory. Interestingly, the best performance is obtained with $\alpha=0.5$, in agreement with the expected overestimation connected with the use of (\ref{deriv-apx}). 

In Fig. \ref{fig2}(c) we explicitly compare the EKF with the optimal inflation for ${\bf P}^f$, ($\rho=0.09$, ${\bf P}^m= \rho {\bf P}^f$), with the EKF implementing the model error treatment through the matrix ${\bf P}^m$ estimated according to (\ref{cov_aprx_3}) and tuned with the optimal values of the scalar coefficient $\alpha=0.5$. The figure displays the estimation error variance as a function of time. 
Note in particular the ability of the filter, using ${\bf P}^m$, to keep estimation error small even in correspondence with the two large deviations experienced by the EKF employing the forecast error covariance inflation.

\subsection{Short Time Augmented Extended Kalman Filter - ST-AEKF }

Let us now turn to the state-augmentation approach. In this case we will assume that model errors arise from mis-specifications of some parameters, so that the theory depicted in Section 2.2 can be used.
This view restricts us to parametric errors, but it also reflects our limited knowledge of the sub-grid scale processes that are only represented through parametrisation schemes for which only a set of parameters is accessible.

A straightforward theory exists for the estimation of the uncertain parameters along with the system's state. The approach, commonly known as state-augmentation \cite{Jazwinski}, consists in defining an augmented dynamical system which allocates, along with the system's state, the model parameters to be estimated. The analysis update is then applied to this new augmented dynamical system. 
Our aim here is to use the state-augmentation approach in conjunction with the deterministic formulation of the model error dynamics. 

The dynamical system (\ref{model-dyn-discrete}), the forecast model, is augmented with the model parameters, as follows:
\begin{equation}
\label{AUG-sys}
{\bf z}^f = \left[ \begin{array}{cc} {\bf x}^f \\  \lam^f  \end{array} \right] = {\mathcal F}{\bf z}^a = 
\left[ \begin{array}{cc} {\mathcal M}{\bf x}^a  \\  {\mathcal F}^\lambda\lam^a  \end{array} \right] 
\end{equation}
where ${\bf z} = ({\bf x}, \lam)$ is the augmented state vector. The augmented dynamical system ${\mathcal F}$ includes the dynamical model for the system's state, ${\mathcal M}$, and a dynamical model for the parameters ${\mathcal F}^{\lambda}$. In the absence of additional information, a persistence model for ${\mathcal F}^{\lambda}$ is usually assumed so that ${\mathcal F}^{\lambda} = {\bf I}$ and $\lam^f_{t_{k+1}}=\lam^a_{t_{k}}$. 
Recently, a temporally smoothed version of the persistence model has been used in the framework of a square root filter \cite{YangDelsole2009}. 
The state-augmented formulation is also successfully applied in the context of the general class of ensemble-based data assimilation procedures \cite{Ruiz2013}. 

By proceeding formally as for Eq. (\ref{dyn-err-solution}) we can write the linearized error evolution for the augmented system, in an arbitrary assimilation interval, $\tau=t_{k+1}-t_{k}$, with initial condition given by the augmented state analysis error, $\delta{\bf z}^a = (\delta{\bf x}^a,\delta\lam^a) = ({\bf x}^a-\ym, \lam^a-\lamm)$:
\begin{equation}
\label{err-AUG}
\delta{\bf z}^f \approx (\delta{\bf x}^f, \delta\lam^f) =  ({\bf M}\delta{\bf x}^a + \int^{t + \tau}_{t} ds{\bf M}_{t,s}\delta\bm\mu^a(s), \delta\lam^a )
\end{equation} 
with $\delta\bm\mu^a = (\frac{\partial {\bf f}}{\partial\lam}\vert_{\lam^a})\delta\lam^a$. The parametric error $\delta\lam^f_{t_{k+1}}=\delta\lam^a_{t_{k}}$ is constant over the assimilation interval in virtue of the assumption ${\mathcal F}^{\lambda} = {\bf I}$. Equation (\ref{err-AUG}) describes, in the linear approximation, the error evolution in the augmented dynamical system (\ref{AUG-sys}). 
The short-time approximation of the error dynamics (\ref{err-AUG}) in the interval $\tau$ reads:
\begin{equation}
\label{err-AUG-st}
\delta{\bf z}^f \approx ({\bf M}\delta{\bf x}^a + \delta\bm\mu^a\tau, \delta\lam^a )
\end{equation} 

As for the standard EKF, by taking the expectation of the product of (\ref{err-AUG}) (or (\ref{err-AUG-st})) with its transpose, we obtain the forecast error covariance matrix, ${\bf P}_z^f$, for the augmented system:
\begin{equation}
\label{PFZ}
{\bf P}^f_z =  <\delta{\bf z}^f\delta{\bf z}^{fT}> = \left( \begin{array}{cc} {\bf P}^f_x & {\bf P}^f_{x\lambda} \\  {\bf P}^{fT}_{x\lambda} & {\bf P}^f_{\lambda}  \end{array} \right)  
\end{equation}
where the $N\times N$ matrix ${\bf P}^f_x $ is the error covariance of the state estimate ${\bf x}^f$, ${\bf P}^f_{\lambda}$ is the $P\times P$ parametric error covariance and ${\bf P}^{f}_{x\lambda}$ the $N\times P$ error correlation matrix between the state vector, ${\bf x}$, and the vector of parameters $\lam$.  
These correlations are essential for the estimation of the parameters. In general one does not have access to a direct measurement of the parameters, and information are only obtained through observations of the system's state. As a consequence, at the analysis step, the estimate of the parameters will be updated only if they correlate with the system's state, that is ${\bf P}^f_{x\lambda}\ne 0$. The gain of information coming from the observations is thus spread out to the full augmented system phase space.

Let us define, in analogy with (\ref{PFZ}), the analysis error covariance matrix for the augmented system:
\begin{equation}
\label{PAZ}
{\bf P}^a_z =  \left( \begin{array}{cc} {\bf P}^a_x & {\bf P}^a_{x\lambda} \\  {\bf P}^{aT}_{x\lambda} & {\bf P}^a_{\lambda}  \end{array} \right)
\end{equation}
where the entries in (\ref{PAZ}) are defined as in (\ref{PFZ}) but refer now to the analysis step after the assimilation of observations. 

By inserting (\ref{err-AUG-st}) into (\ref{PFZ}), and taking the expectation, we obtain the forecast error covariance matrix in the linear and short-time approximation:
\begin{equation}
{\bf P}^f_x = {\bf M} <\delta{\bf x}^a{\delta{\bf x}^a}^{T}> {\bf M}^T + <\delta\bm\mu^a{\delta\bm\mu^a}^{T}>\tau^2 + \nonumber 
\end{equation} 
\begin{equation}
[{\bf M}<\delta{\bf x}^a{\delta\bm\mu^a}^{T}> + <\delta\bm\mu^a{\delta{\bf x}^a}^{T}>{\bf M}^T]\tau \nonumber
\end{equation} 
\begin{equation}
\label{PFX}
= {\bf MP}^a_x{\bf M}^T + {\bf Q}^a\tau^2 + [{\bf M}<\delta{\bf x}^a{\delta\bm\mu^a}^{T}> + <\delta\bm\mu^a{\delta{\bf x}^a}^{T}>{\bf M}^T]\tau
\end{equation} 
\begin{equation}
\label{PFL}
{\bf P}^f_\lam = <\delta\lam^a \delta\lam^{aT}>
\end{equation} 
\begin{equation}
\label{PFLX}
{\bf P}^f_{x\lambda} = {\bf M}<\delta{\bf x}^a \delta\lam^{aT}> + <\delta\bm\mu^a \delta\lam^{aT} > \tau
\end{equation} 
Note that (\ref{PFX}) is equivalent to (\ref{PF-DETER}), except that now the correlations between the initial condition and the model error are maintained (last two terms on the r.h.s. of  (\ref{PFX})), and ${\bf P}_x^a$ and ${\bf Q}^a$ replace ${\bf P}^a$ and ${\bf Q}$.
Nevertheless, in contrast to the ST-EKF where ${\bf Q}$ is estimated statistically and then kept fixed, in the ST-AEKF ${\bf Q}^a$ is estimated online using the observations.  

The information on the uncertainty in the model parameters is embedded in the error covariance ${\bf P}_\lam^a$, a by-product of the assimilation. Using the definition of $\delta\bm\mu^a$ and (\ref{PFL}), the matrix ${\bf Q}^a$ can be rewritten as:
\begin{equation}
{\bf Q}^a = <\delta\bm\mu^a \delta\bm\mu^{aT}> = \nonumber
\end{equation}
\begin{equation}
\label{QAUG}
< \left(\frac{\partial {\bf f}}{\partial\lam}\vert_{\lam^a}\right)\delta\lam^a \delta\lam^{aT} \left(\frac{\partial {\bf f}}{\partial\lam}\vert_{\lam^a}\right)^T > \approx \left(\frac{\partial {\bf f}}{\partial\lam}\vert_{\lam^a}\right) {\bf P}^a_{\lam} \left(\frac{\partial {\bf f}}{\partial\lam}\vert_{\lam^a}\right)^T
\end{equation}
Similarly, the correlation terms in (\ref{PFX}) can be written according to:
\begin{equation}
[{\bf M}<\delta{\bf x}^a{\delta\bm\mu^a}^{T}> + <\delta\bm\mu^a{\delta{\bf x}^a}^{T}>{\bf M}^T]\tau = \nonumber
\end{equation}
\begin{equation}
[{\bf M}<\delta{\bf x}^a{\delta\lam^a}^{T} \left(\frac{\partial {\bf f}}{\partial\lam}\vert_{\lam^a}\right)^T> + < \left(\frac{\partial {\bf f}}{\partial\lam}\vert_{\lam^a}\right)\delta\lam^a{\delta{\bf x}^a}^{T}>{\bf M}^T]\tau \approx \nonumber
\end{equation}
\begin{equation}
\label{Corr-aug}
[{\bf M}{\bf P}^a_{{\bf x}\lam}\left(\frac{\partial {\bf f}}{\partial\lam}\vert_{\lam^a}\right)^T + \left(\frac{\partial {\bf f}}{\partial\lam}\vert_{\lam^a}\right){{\bf P}^a_{{\bf x}\lam}}^{T}{\bf M}^T]\tau 
\end{equation}
Using (\ref{QAUG}) and (\ref{Corr-aug}) in (\ref{PFX}), the forecast state error covariance ${\bf P}^f_{x}$ can be written in terms of the state-augmented analysis error covariance matrix at the last observation time, according to:
\begin{equation}
{\bf P}^f_x \approx {\bf MP}^a_x{\bf M}^T +  \left(\frac{\partial {\bf f}}{\partial\lam}\vert_{\lam^a}\right) {\bf P}^a_{\lam} \left(\frac{\partial {\bf f}}{\partial\lam}\vert_{\lam^a}\right)^T\tau^2 +  \nonumber
\end{equation}
\begin{equation}
\label{PFX_app}
[{\bf M}{\bf P}^a_{{\bf x}\lam}\left(\frac{\partial {\bf f}}{\partial\lam}\vert_{\lam^a}\right)^T + \left(\frac{\partial {\bf f}}{\partial\lam}\vert_{\lam^a}\right){{\bf P}^a_{{\bf x}\lam}}^{T}{\bf M}^T]\tau
\end{equation}
The three terms in (\ref{PFX_app}) represent the contribution to the forecast state error covariance coming from the analysis error covariance in the system's state, in the parameters and in their correlation respectively. 

By making use of the definition of the model error vector $\delta\bm\mu^a$ in (\ref{PFLX}), the forecast error correlation matrix ${\bf P}_{x\lam}^f$ becomes:
\begin{equation}
\label{PFLX_app}
{\bf P}^f_{x\lam} \approx {\bf M}{\bf P}^a_{x\lam} + \left(\frac{\partial {\bf f}}{\partial\lam}\vert_{\lam^a}\right) {\bf P}^a_{\lam} \tau
\end{equation}
Expressions (\ref{PFL}), (\ref{PFX_app}) and (\ref{PFLX_app}) can be compacted into a single expression:
\begin{equation}
\label{forstep_ST-AEKF}
{\bf P}^f_z = {\bf C}{\bf P}^a_z{\bf C}^T
\end{equation}
with ${\bf C}$ being the ST-AEKF forward operator defined as:
\begin{equation}
\label{C_ST-AEKF}
{\bf C} =  \left( \begin{array}{cc} {\bf M} & \frac{\partial {\bf f}}{\partial\lam}\vert_{\lam^a}\tau  \\  0  & {\bf I}_{P}  \end{array} \right)
\end{equation}
where ${\bf I}_{P}$ is the $P\times P$ identity matrix.

The short-time bias equation (\ref{errore-medio}) is used to estimate the bias in the state forecast, ${\bf x}^f$, due to parametric error, in analogy with the ST-EKF. This estimate is made online using the last innovation of the parameter vector. Assuming furthermore that the forecast of the parameter is unbiased, the bias in the state augmented forecast at time $t_{k+1}$ reads:
\begin{equation}
\label{bias_aug}
{\bf b}^m_z =  \left( \begin{array}{c} {\bf b}_x  \\  {\bf b}_{\lam}  \end{array} \right) = \left( \begin{array}{c} \frac{\partial {\bf f}}{\partial\lam}\vert_{\lam^a}(\lam^a_{t_{k}}-\lam^f_{t_{k}})\tau \\ 0 \end{array} \right)
\end{equation}
The bias ${\bf b}^m_z$ is then removed from the forecast field before the latter is used in the analysis update, that is ${\tilde {\bf z}}^f = {\bf z}^f - {\bf b}^m_z$, where ${\tilde {\bf z}}^f$ is the unbiased state augmented forecast. 

As for the standard EKF, we need the observation operator linking the model to the observed variables. An augmented observation operator is introduced, ${\mathcal H}_z = [{\mathcal H}\quad 0]$ with ${\mathcal H}$ as in (\ref{obs}), and its linearization, ${\bf H}_z$ is now a $M\times (N+P)$ matrix in which the last $P$ columns contain zeros; the rank deficiency in ${\mathcal H}$ reflects the lack of direct observations of the model parameters.

The augmented state and covariance update complete the algorithm:
\begin{equation}
\label{Z_update}
{\bf z}^{a}=\left[{\bf I}_z -{\bf K}_z{\bf H}_z\right]{\tilde {\bf z}}^{f} + {\bf K}_z{\bf y}^{o}
\end{equation}
\begin{equation}
\label{AN_COV_UPD_AUG}
{\bf P}^a_z = [{\bf I}_z - {\bf K}_z{\bf H}_z]{\bf P}^f_z
\end{equation}
where the vector of observations ${\bf y}^{o}$ is the same as in the standard EKF while ${\bf I}_z$ is now the $(N+P)\times(N+P)$ identity matrix. The augmented gain matrix ${\bf K}_z$ is defined accordingly:
\begin{equation}
\label{gain matrix-aug}
{\bf K}_z={\bf P}^{f}_z{\bf H}^T_z\left[{\bf H}_z{\bf P}^{f}_z{\bf H}^T_z + {\bf R}\right]^{-1}
\end{equation}
but it is now a $(I+P)\times M$ matrix.

Equations (\ref{AUG-sys}) - (\ref{forstep_ST-AEKF}) for the forecast step, and (\ref{Z_update}) - (\ref{gain matrix-aug}) for the analysis update define the ST-AEKF. The algorithm is closed and self consistent meaning that, once it has been initialized, it does not need any external information (such as statistically estimated error covariances) and the state, the parameters and the associated error covariances are all estimated online using the observations. 

The ST-AEKF is a short-time approximation of the classical augmented EKF, the AEKF \cite{Jazwinski}. 
In essence, the approximation consists of the use of an analytic expression for the evolution of the model error component of the forecast error covariance. This evolution law, quadratic for short-time, reflects a generic and intrinsic feature of the model error dynamics, connected to the model sensitivity, to perturbed parameters and to the degree of dynamical instability. It does not depend on the specific numerical integration scheme adopted for the evolution of the model state. The state error covariance, ${\bf P}_x^f$, in the ST-AEKF, is evolved as in the standard AEKF: the propagator ${\bf M}$ is the product of the individual ${\bf M}_i$ relative to each time-step within the assimilation interval. The difference between the two algorithms is in the time propagation of the forecast error covariance associated with the misspecification of the parameters, ${\bf P}_{x\lambda}$. In the ST-AEKF this is reduced to the evaluation of the off diagonal term in the operator ${\bf C}$. This term replaces the full linearization of the model equations with respect to the estimated parameters, required by the AEKF. 
In this latter case the model equations are linearized with respect to the augmented state, $({\bf x},\lam)$, giving rise to an augmented tangent linear model, ${\bf M}_{\bf z}$. This linearization can be particularly involved \cite{Kondrashov2007}, especially in the case of implicit or semi-implicit integration schemes such as those often used in NWP applications \cite{Kalnay2002}. The propagator relative to the entire assimilation interval is then given by the product of the individual augmented tangent linear propagator over the single time-steps. As a consequence the cost of evolving the model error covariance in the AEKF grows with the assimilation interval.      
In the ST-AEKF, the use of the short-time approximation within the assimilation interval makes straightforward the implementation of the parameter estimation in the context of a pre-existing EKF, without the need to use an augmented tangent linear model during the data assimilation interval. It reduces the computational cost with respect to the AEKF, because the propagation of the model error component does not depend on the length of the assimilation interval. 
Nevertheless the simplifications in the setup and the reduction in the computational cost are obtained at the price of a decrease in the accuracy with respect to the AEKF.  
The degree of dynamical instabilities along with the length of the assimilation interval, are the key factors affecting the accuracy of the ST-AEKF.

\subsubsection{Numerical Results with ST-EKF and ST-AEKF}

Numerical experiments are carried out with two different models. 
OSSEs are performed first using the Lorenz '96 \cite{Lorenz1996} model used in Sect. 3.1.1, but in its one-scale version given by:
\begin{equation}
\label{model-expr}
\frac{dx_i}{dt}=\alpha(x_{i+1}-x_{i-2})x_{i-1} -  \beta x_i + F, \qquad  i=\{1,...,36\}
\end{equation}
where the parameter associated with the advection, $\alpha$, linear dissipation, $\beta$ and the forcing $F$, are written explicitly.  
As for the experiments described in Sect. 3.1.1, the numerical integration are performed using a fourth-order Runge-Kutta scheme with a time step of 0.0083 units, corresponding to 1 hour of simulated time.

The reference trajectory, representing the true evolution we intend to estimate, is given by a solution of (\ref{model-expr}) with parameters $\lam^{tr} = (F^{tr},\alpha^{tr},\beta^{tr}) = (8,1,1)$; with this choice the model behaves chaotically. A network of $M=18$ regularly spaced noisy observations is simulated by sampling the reference true trajectory and adding a Gaussian random observation error whose variance is set to $5\%$ of the system's climate variance. Model error is simulated by perturbing $F$, $\alpha$ and $\beta$ with respect to their reference true values. 
Gaussian samples of $100$ states and model parameters are used to initialize assimilation cycles lasting for $1$ year. In all the experiments the initial condition error variance is set to $20\%$ of the system's climate variance. The model parameters are sampled from a Gaussian distribution with mean equal to $\lam^{tr}$ and standard deviation $\sigma_{\lambda}=25\%$ of $\lam^{tr}$.

We compare four configurations of the EKF: (1) standard EKF without model error treatment, (2) standard EKF using a perfect model, (3) ST-EKF (Sect. 3.1), and (4) ST-AEKF (Sect. 3.2). 
Recall that in the ST-EKF, model error bias and covariance are estimated according to (\ref{bias-PM}) and (\ref{PPM}) with $<\delta\bm\mu_o>$ and ${\bf Q}$ evaluated on a statistical basis before the assimilation experiments. 
The expectation of $\delta\bm\mu_o$ is estimated through:
\begin{equation}
\label{est_mean}
<\delta\bm\mu_o> = <<\frac{\partial {\bf f}}{\partial\lam}\vert_{\lam}(\lam-\lam^{tr})>>
\end{equation}
and is then used in (\ref{bias-PM}) and (\ref{PPM}).
In (\ref{est_mean}) the averages are taken over the same Gaussian sample of initial conditions and parameters used to initialize the data assimilation experiments, using the actual value of the parameter, $\lam^{tr}$, as the reference. This idealized procedure has been chosen to give the ST-EKF the best-possible statistical estimate of the model error in view of its comparison with the more sophisticated ST-AEKF. 

\begin{figure}[h]
\begin{center}
\includegraphics[height=6.4cm,width=9.5cm]{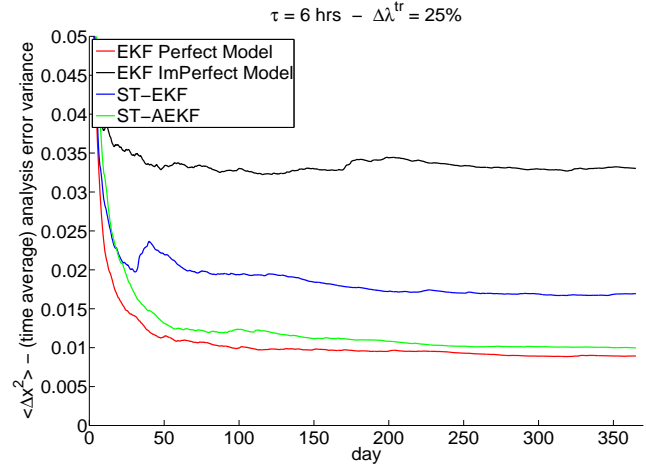}
\end{center}
\caption{Time averaged analysis error variance as a function of time. Standard EKF without model error treatment (black), standard EKF with perfect model (red), ST-EKF (blue) and ST-AEKF (green). The error variance is normalized with respect to the system's climate variance.}
\label{FIG3}
\end{figure}

Figure \ref{FIG3} shows the analysis error variance as a function of time for the four experiments of one year long; the assimilation interval is $\tau=6$ hours. The errors are spatio-temporal average over the ensemble of $100$ experiments and over the model domain, and normalized with the system's climate variance.  
The figure clearly shows the advantage of incorporating a model error treatment: the average error of the ST-EKF is almost half of the corresponding to the standard EKF without model error treatment. However using the ST-AEKF the error is further reduced and attains a level very close to the perfect model case. 

\begin{figure}[h]
\begin{center}
\includegraphics[height=6.4cm,width=9.5cm]{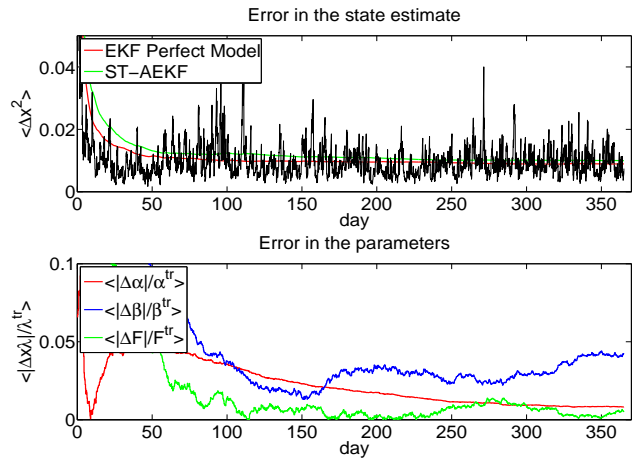}
\end{center}
\caption{Top Panel - Time averaged analysis error variance as a function of time: standard EKF with perfect model (red) and ST-AEKF (green); time series of the ST-AEKF (black). Bottom Panel - Absolute parametric error of the ST-AEKF, relative to the true value $\lam^{tr}$. The error variance is normalized with respect to the system's climate variance.}
\label{FIG4}
\end{figure}

The benefit of incorporating a parameter estimation procedure in the ST-AEKF is displayed in Fig. \ref{FIG4} that shows the time mean analysis error variance for the EKF with perfect model and the ST-AEKF (top panel), along with the relative parametric errors as a function of time (bottom panel). The time series of the ST-AEKF analysis error variance is also superimposed to the time-averages in the top panel. Figure \ref{FIG4} reveals that the ST-AEKF is successful in recovering the true parameters. This reconstruction is very effective for the forcing, $F$, and the advection, $\alpha$, and at a lesser extent for the dissipation, $\beta$.       
The ability of the ST-AEKF to efficiently exploit the observations of the system's state to estimate an uncertain parameter, either multiplicative or additive, is evident. Given that the innovation in the parameter, obtained via Eq. (\ref{Z_update}), is proportional to the cross-covariance forecast error, ${\bf P}_{x\lambda}^f$, the accuracy of the parameter estimation revealed by Fig. \ref{FIG4} turns out to be an indication of the quality of the short-time approximation, (\ref{PFLX_app}), on which the estimate of ${\bf P}_{x\lambda}^f$ is based. 

\begin{figure}[h]
\begin{center}
\includegraphics[height=6.5cm,width=13cm]{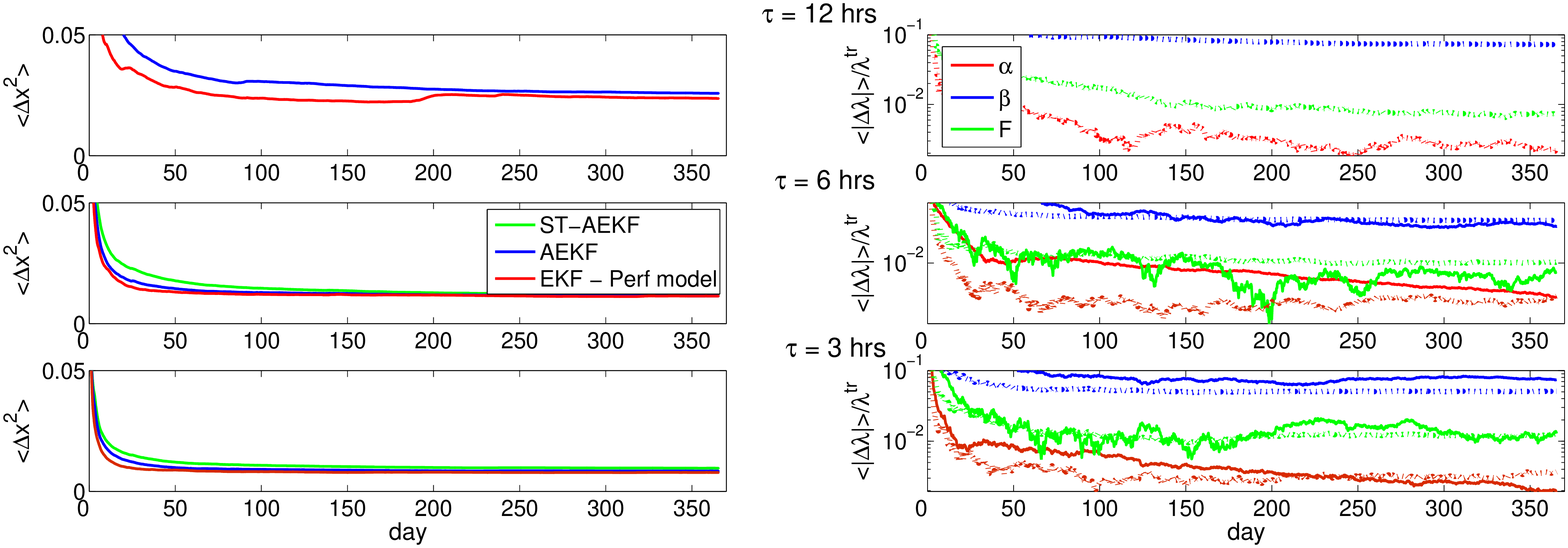}
\end{center}
\caption{
Left column: Running mean of the quadratic state estimation error as a function of time; EKF with prefect model (red), ST-AEKF (green) and AEKF (blue). Right column: absolute value of the parametric error as a function of time for $\alpha$ (red), $\beta$ (blue) and $F$ (green), for ST-AEKF (solid lines) and AEKF (dotted lines).From top to bottom $\tau = 12$, $6$ and $3$ hours respectively. The errors are averaged over an ensemble of $100$ experiments and $\sigma_{\lambda}=25\%\lambda^{tr}$.}
\label{FIG5}
\end{figure}

Figure \ref{FIG5} focuses on the comparison between the ST-AEKF and the standard AEKF. 
The experiments are carried out for $\tau=3$, $6$ and $12$ hours and with $\sigma_{\lambda}=25\%\lam^{tr}$. As above, the results are averaged over an ensemble of $100$ experiments, and the observation error variance is $5\%$ of the system's climate variance. The left panels display the quadratic estimation error, while the parametric error is given in the panels in the right column; note that the logarithm scale is used in the y-axis. The estimation error relative to the EKF with a perfect model is also displayed for reference.  

We see that as expected the AEKF has a superior skill than the ST-AEKF but for $\tau=3$ or $6$ hours their performances are very similar. The AEKF shows a marked rapidity to reach convergence but the asymptotic error level attained by the two filters are practically indistinguishable. On the other hand for $\tau=12$ hours the ST-AEKF diverges whereas the AEKF is able to control error growth and maintain the estimation error to a low level.   
We first observe that in all but one cases the parametric error in the experiments with the AEKF is lower than for the ST-AEKF, in agreement with the observed lower state estimation error. Anyhow when $\tau=6$ or $3$ hours, the asymptotic parametric errors of the two filters are very similar, a remarkable result considering the approximate evolution law used in the ST-AEKF. An important difference is the extreme variability observed in the parametric error with the ST-AEKF as compared to the smoothness of the corresponding solutions with the AEKF. Note also that when $\tau=6$ hours the ST-AEKF reduces the error in the forcing more than the AEKF but the error curves are subject to very large fluctuations. The dissipation, $\beta$, appears as the most difficult parameter to be estimated in agreement with what observed in Fig. \ref{FIG4}. In summary, Fig. \ref{FIG5} suggests that the ST-AEKF may represent a suitable and efficient alternative to the full AEKF when the assimilation interval does not exceed the time range of validity of the approximation on which the ST-AEKF is based. The results indicate that this limit is between $6$ and $12$ hours given that the ST-AEKF diverges when $\tau=12$ hours. According to the theory outlined in \cite{Nicolis2003}, the short-time regime is related to the inverse of twice the largest (in absolute value) Lyapunov exponent of the system. In the Lorenz system (\ref{model-expr}) the largest Lyapunov exponent turns out to be the most negative one, equal to $0.97$ day$^{-1}$, so that the duration of the short-time regime is estimated to be about $12$ hours, in qualitative agreement with the performance of the ST-AEKF. Finally note that the slight deterioration in the filter accuracy is compensated by a reduction in both the computational and implementation costs with respect to the AEKF. 

The second model under consideration is an offline version of the operational soil model, Interactions between Surface, Biosphere,and Atmosphere (ISBA) \cite{NoilhanMahfouf1996}. In the experiments that follow, ST-AEKF has been implemented in the presence of parametric errors in the Leaf Area Index (LAI) and in the Albedo; more details, along with the case of other land surface parameters, can be found in \cite{C_ISBA12}. 
OSSEs are performed using the two-layers version of ISBA which describes the evolution of soil temperature and moisture contents; the model is available within a surface externalized platform (SLDAS; \cite{Mahfouf2007}).
The state vector, ${\bf x}= (T_s,T_2,w_g,w_2)$, contains the surface and deep soil temperatures  $T_s$ and $T_2$ and the corresponding water contents $w_g$ and $w_2$. The vector $\lam$ is taken to represent the set of model parameters.
A detailed description of ISBA can be found in \cite{NoilhanMahfouf1996}.

The forcing data are the same for the truth and the assimilation solutions. They consist of 1-hourly air temperature,
specific humidity, atmospheric pressure, incoming global radiation, incoming long-wave radiation, precipitation rate and wind speed relative to the ten summers in the decade 1990-1999 extract from  ECMWF Re-analysis ERA40 and
then dynamically down-scaled to 10 km horizontal resolution over Belgium \cite{Hamdi2012}.
The fields are then temporally
interpolated to get data consistent with the time resolution of the integration scheme of ISBA (300 s).
In this study ISBA is run in one offline single column mode for a 90 day period, and the forcing parameters are those
relative to the grid point closest to Brussels. An one-point soil model has been also used by \cite{Orescanin2009},
for parameter estimation using an ensemble based assimilation algorithm.

The simulated observations are $T_{2m}$ and $RH_{2m}$, interpolated between the forcing level ($\approx$20 m) and the surface with the Geleyn's interpolation scheme (\cite{Geleyn1988}), at $00$, $06$, $12$ and $18$ UTC. The assimilation interval is $\tau=6$ hours, while the observational noise is drawn from a Gaussian, $\mathcal{N}$$(0,{\bf R})$, with zero-mean and covariance given by the diagonal matrix ${\bf R}$ with elements: $diag({\bf R}) = (\sigma^2_{T_{2m}},\sigma^2_{w_{2m}}) = (1K^2,10^{-2})$. As explained in \cite{Mahfouf2009}, the observation operator$\mathcal{H}$, relating the state vector to the observation includes the model integration.
The initial ${\bf P}^a$ and ${\bf P}^m$ required by the EKF are set as diagonal with elements
$diag({\bf P}^a) = (\sigma^2_{T_s},\sigma^2_{T_2},\sigma^2_{w_g},\sigma^2_{w_2})= (1 K^2,1 K^2,10^{-2},10^{-2})$
\footnote{{\scriptsize The values of $\sigma_{w_g}$ and $\sigma_{w_2}$ are expressed as soil wetness index $SWI = (w - w_{wilt})/(w_{fc} - w_{wilt})$ where $w_{fc}$ is the volumetric field capacity and $w_{wilt}$ is the wilting point. }}
and $diag({\bf P}^m) = (\sigma^2_{T_s},\sigma^2_{T_2},\sigma^2_{w_g},\sigma^2_{w_2}) = (25\times10^{-2} K^2,25\times10^{-2} K^2,4\times10^{-4},4\times10^{-4})$ \cite{Mahfouf2007}.

Parametric errors is introduced by perturbing simultaneously the LAI and the albedo. These parameters strongly influence the surface energy balance budget and partitioning, which in turn regulate the circulation patterns and modify the hydrological processes.
For each summer in the period 1990-1999, a reference trajectory is generated by integrating the model with
LAI = 1 $m^{2}/m^{2}$ and albedo = 0.2. Around each of these trajectories, Gaussian samples of $100$ initial
conditions and uncertain parameters are used to initialize the assimilation cycles. The initial conditions are sampled
from a distribution with standard deviation $(\sigma_{T_s},\sigma_{T_2},\sigma_{w_g},\sigma_{w_2})=(5 K,5 K,1,1)$, whereas LAI and the albedo are sampled with standard deviations, $\sigma_{LAI}= 0.5$ $m^2/m^2$ and $\sigma_{albedo} = 0.05$ respectively (\cite{Ghilain2011}).
The initial ${\bf P}^a_{\lam}$ in the ST-AEKF read ${\bf P}^a_{LAI}=1$ $(m^2/m^2)^2$ and ${\bf P}^a_{albedo}=10^{-4}$, ${\bf P}^a_{{\bf x}}$ is taken as in the EKF while ${\bf P}^a_{{\bf x},\lam}$ is initially set to zero.

\begin{figure}[h]
\begin{center}
\includegraphics[height=8cm,width=12cm]{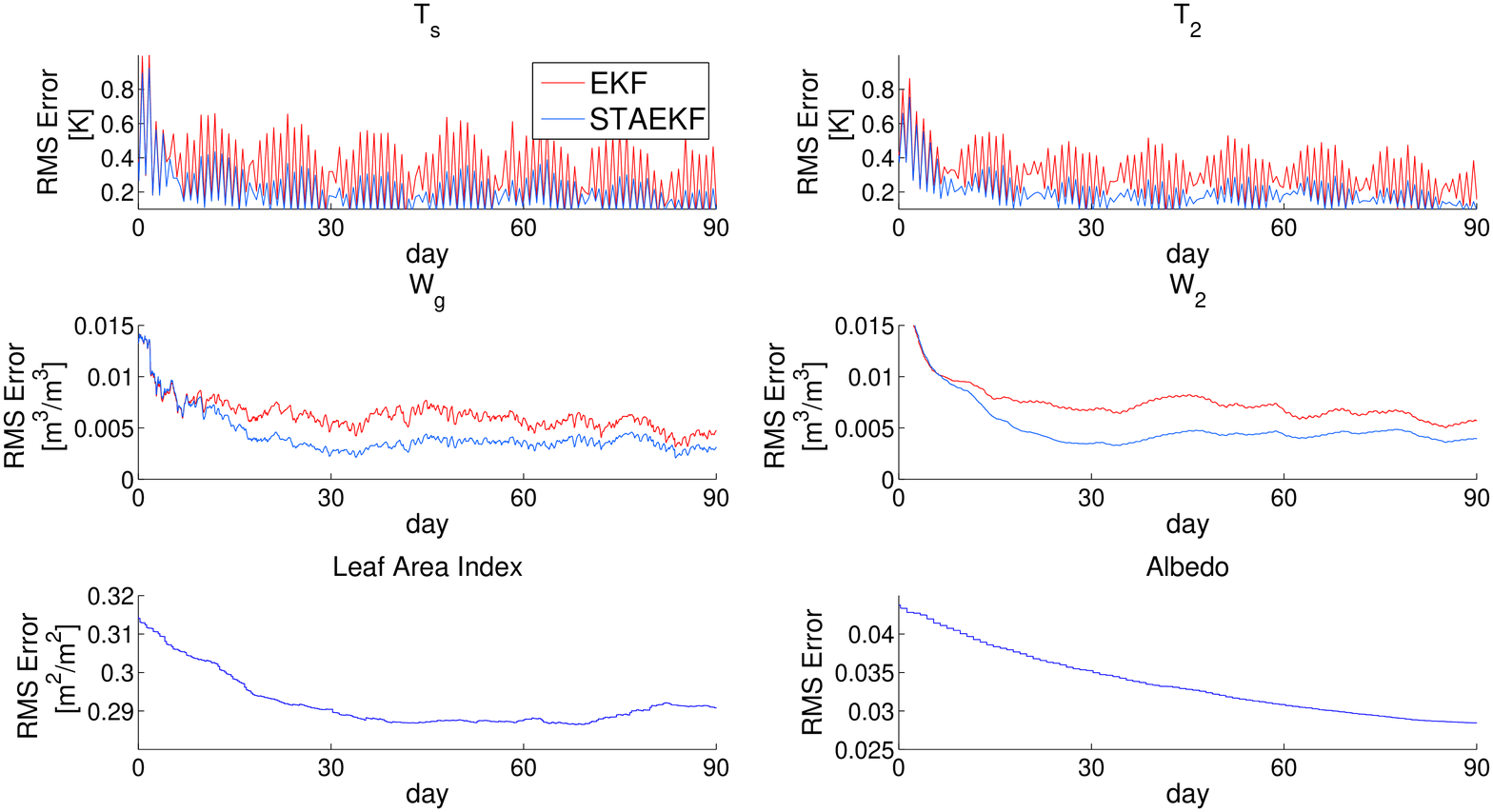}
\end{center}
\caption{RMS estimation error in the four state variables for the EKF (red) and the STAEKF (blue). The RMS error in the estimate of $LAI$ and Albedo relative to the ST-AEKF are shown in the bottom-left/right panels respectively. From \cite{C_ISBA12}.}
\label{FIG6}
\end{figure}

Results are summarized in Fig. \ref{FIG6} which shows the RMS Error in the four state variable for the EKF and ST-AEKF, along with the RMS Error in LAI and Albedo for the ST-AEKF. The progressive parametric error reduction achieved with the ST-AEKF is reflected by the systematically lower estimation error in the soil temperature and water content. At the very initial times, on the order of one week, EKF and ST-AEKF have an indistinguishable skill. However, as soon as, the state-parameter error correlations in the ST-AEKF augmented forecast error matrix become mature, the improvement of the ST-AEKF becomes apparent and it lasts for the entire duration of the experiment.         
By reducing the parametric error a better guess for the system state can be obtained and this in turn improves the analysis field and again the accuracy of the parameter estimate. Moreover, given that this feature is incorporated using the short-time formulation \cite{CV11a}, the additional computational cost with respect to the standard EKF is almost negligible.

\section{Deterministic model error treatment in variational data assimilation \label{sec:VAR}}

 Variational assimilation attempts to solve the smoothing problem of a simultaneous fit to all observations distributed within a given interval of interest. We suppose therefore that $I$ measurements, (\ref{obs}), are collected at the discrete times, $(t_1, t_2, ..., t_I) $, within a reference time interval $T$.
An priori estimation, ${\bf x}^b$, of the model initial condition is supposed to be available. This is usually referred to as the background state, and:
\begin{equation}
\label{back}
{\bf x}_0={\bf x}_b + {\bf\epsilon}_b 
\end{equation}
where $\bf\epsilon_b$ represents the background error.

We search for the trajectory that, on the basis of the background field and according to some specified criteria, best fits the observations over the reference period $T$. Besides the observations and the background, the model dynamics itself represents a source of additional information to be exploited in the state estimate. The model is not perfect, and we assume that an additive error affects the model prediction in the form:
\begin{equation}
\label{mod-usato}
{\bf x}(t) = \mathcal{M}({\bf x}_0) + \delta{\bf x}^m(t) 
\end{equation}

Assuming furthermore that all errors are Gaussian and do not correlate with each other, the quadratic penalty functional, combining all the information, takes the form \cite{Jazwinski}:
\begin{equation}  
2J=\int_{0}^{T}\int_{0}^{T} (\delta{\bf x}^m(t^{'}))^T({\bf P}^m)^{-1}_{t^{'}t^{''}}(\delta{\bf x}^m(t^{''})) dt^{'}dt^{''}  + \nonumber
\end{equation}
\begin{equation}  
\label{CF}
\sum_{k=1}^{I}(\bf\epsilon^o_k)^T{\bf R}_k^{-1}(\bf\epsilon^o_k) + \bf\epsilon_b^T{\bf B}^{-1}\bf\epsilon_b
\end{equation}
The weighting matrices ${\bf P}_{t^{'}t^{''}} = {\bf P}(t^{'},t^{''})$, ${\bf R}_k$ and ${\bf B}$ have to be regarded as a measure of our confidence in the model, in the observations and in the background field, respectively. In this Gaussian formulation these weights can be chosen to reflect the relevant moments of the corresponding Gaussian error distributions. The best-fit is defined as the solution, $\hat{\bf x}(t)$, minimizing the cost-function $J$ over the interval $T$. It is known that, under the aforementioned hypothesis of Gaussian errors, $\hat{{\bf x}}(t)$ corresponds to the maximum likelihood solution and $J$ can be used to define a multivariate distribution of ${\bf x}(t)$ \cite{Jazwinski}. Note that, in order to minimize $J$ all errors have to be explicitly written as a function of the trajectory ${\bf x}(t)$. 

The variational problem defined by (\ref{CF}) is usually referred to as {\it weak-constraint} given that the model dynamics is affected by errors \cite{Sasaki1970}. An important particular case is the {\it strong-constraint} variational assimilation in which the model is assumed to be perfect, that is $\delta{\bf x}^m = 0$, \cite{DimetTalag1986, LewDer1985}. 
In this case the model-error related term disappears and the cost-function reads:
\begin{equation}
\label{CF-strong}
2J_{strong}= \sum_{k=1}^{I}(\bf\epsilon^o_k)^T{\bf R}_k^{-1}(\bf\epsilon^o_k) + \bf\epsilon_b^T{\bf B}^{-1}\bf\epsilon_b 
\end{equation}

The calculus of variations can be used to find the extremum of (\ref{CF}) (or (\ref{CF-strong})) and leads to the corresponding Euler-Lagrange equations \cite{DimetTalag1986, Bennett1992}. In the strong-constraint case, the requirement that the solution has to follow the dynamics exactly is satisfied by appending to (\ref{CF-strong}) the model equations as a constraint by using a proper Lagrange multiplier field. However the size and complexity of the typical NWP problems is such that the Euler-Lagrange equations cannot be practically solved unless drastic approximations are introduced. When the dynamics is linear and the amount of observations is not very large, the Euler-Lagrange equations can be efficiently solved with the method of representers \cite{Bennett1992}. An extension of this approach to nonlinear dynamics has been proposed in \cite{UboKam2000}. Nevertheless, the representers method is far from being applicable for realistic high dimensional problems, like the numerical weather prediction and an attractive alternative is represented by the descent methods which makes use of the gradient vector of the cost-function in an iterative minimization procedure \cite{TalagCourt1987}. This latter approach is used in most of the operational NWP centers which employ variational assimilation. Note that in the cost-functions (\ref{CF}) model error is allowed to be correlated in time, and gives up the double integral in the first r.h.s. term. If model error is assumed to be a random uncorrelated noise, only covariances have to be taken into account and the double integral reduces to a single integral (to a single summation in the discrete times case). 

The search for the best-fit trajectory by minimizing the associated cost-function requires the specification of the weighting matrices. The estimation of the matrices ${\bf P}^m_{t^{'}t^{''}}$ is particularly difficult in realistic NWP applications due to the large size of the typical models currently in use. Therefore it becomes crucial to define approaches for modeling the matrices ${\bf P}^m_{t^{'}t^{''}}$ and reduce the number of parameters required for their estimation.
We will show below how the deterministic, and short-time, model error formulation described in Sect. 2.2 can be used to derive ${\bf P}^m_{t^{'}t^{''}}$ 

We make the conjecture that, as long as the errors in the initial condition and in the model parameters are small, the second rhs term of (\ref{dyn-err-solution}), $\delta{\bf x}^{m}(t)=\int^{t}_{t_{0}} d\tau{\bf M}_{t,\tau}\delta\bm\mu(\tau)$ can be used to estimate the model error entering the weak-constraint cost-function, and the corresponding correlation matrices ${\bf P}^m(t^{'},t^{''})$. In this case, the model error dependence on the model state, induces the dependence of model error correlation on the correlation time scale of the model variables themselves.
By taking the expectation of the product of the second rhs term of (\ref{dyn-err-solution}) by itself, over an ensemble of realizations around a specific trajectory, we obtain an equation for the model error correlation matrix:
\begin{equation}
\label{PMOD}
{\bf P}^m(t^{'},t^{''}) =  \int^{t^{'}}_{t_{0}} d\tau \int^{t^{''}}_{t_{0}} d\tau^{'} {\bf M}_{t^{'},\tau}< \delta\bm\mu(\tau)\delta\bm\mu(\tau^{'})^T> {\bf M}^T_{t^{''},\tau^{'}}
\end{equation}
The integral equation (\ref{PMOD}) gives the model error correlation between times $t^{'}$ and $t^{''}$. In this form, Eq. (\ref{PMOD}) is of little practical use for any realistic non-linear systems. A suitable expression can be obtained by considering its short-time approximation through a Taylor expansion around $(t^{'},t^{''}) = (t_{0},t_{0})$. It can be shown (\cite{CV10}) that the first non-trivial order is quadratic and reads:
\begin{equation}
\label{PMOD-app}
{\bf P}(t^{'},t^{''}) \approx < \delta\bm\mu_0\delta\bm\mu_0^T> (t^{'}-t_0)(t^{''}-t_0) 
\end{equation}
Equation (\ref{PMOD-app}) states that the model error correlation between two arbitrary times, $t^{'}$ and $t^{''}$, within the short-time regime, is equal to the model error covariance at the origin, $< \delta\bm\mu_0\delta\bm\mu_0^T>$, multiplied by the product of the two time intervals. Naturally the accuracy of this approximation is connected on the one hand to the length of the reference time period, on the other to the accuracy of the knowledge about the error in the parameters needed to estimate  $< \delta\bm\mu_0\delta\bm\mu_0^T>$. 
We propose to use the short-time law (\ref{PMOD-app}) as an estimate of the model error correlations in the variational assimilation. The resulting algorithm is hereafter referred to as Short-Time-Weak-Constraint-4DVar (ST-w4DVar). Besides the fact of being a short-time approximation, (\ref{PMOD-app}) is based on the hypothesis of linear error dynamics. To highlight advantages and drawbacks of its application, we explicitly compare ST-w4DVar with other formulations.

\subsection{Numerical Results with ST-w4DVar}

The analysis is carried out in the context of two systems of increasing complexity. We first deal with a very simple example of scalar dynamics which is fully integrable. The variational problem is solved with the technique of representers. The simplicity of the dynamics allows us to explicitly solve (\ref{PMOD}) and use it to estimate the model error correlations. This "full weak-constraint" formulation of the 4DVar is evaluated and compared with the ST-w4DVar employing the short-time evolution law (\ref{PMOD-app}). In addition, a comparison is made with the widely used strong-constraint 4DVar in which the model is considered as perfect.
In the last part of the Section we extend the analysis to an idealized nonlinear chaotic system. In this case the minimization is made by using an iterative descent method which makes use of the cost-function gradient. In this nonlinear context ST-w4DVar is compared to the strong-constraint and to a weak-constraint 4DVar in which model error is treated as a random uncorrelated noise as it is often assumed in realistic applications.

Let us consider the simple scalar dynamics:
\begin{equation}
\label{sys1}
x(t)=x_0e^{\lambda^{tr} t}
\end{equation}
with $\lambda^{tr} > 0$, as our reference. 

Suppose that $I$ noisy observations of the state variable are available at the discrete times $t_{k}\in[0,T]$, $1\le k \le I$:
\begin{equation}
y^o_k = x_k + \epsilon_k^o \nonumber
\end{equation}
$\epsilon_k^o$ being an additive random noise with variance $\sigma_o^2(t_k)=\sigma_o^2$, $1\le k \le I$, and that a background estimate, $x_b$, of the initial condition, $x_0$, is at our disposal:
\begin{equation}
x_0 = x_b + \epsilon_b \nonumber
\end{equation}
with $\epsilon_b$ being the background error with variance $\sigma_b^2$. 
We assume the model is given by: 
\begin{equation}
x(t) = x_0e^{\lambda t}. \nonumber
\end{equation}
We seek for a solution minimizing simultaneously the error associated with all these information sources. The quadratic cost function can be written in this case as:
\begin{equation}
2J(x) = \int_{0}^{T}\int_{0}^{T}  (x(t^{'})-x_0e^{\lambda t^{'}})p^{-2}_{t^{'}t^{''}}( x(t^{''})-x_0e^{\lambda t^{''}}   )  dt^{'}dt^{''}  + \nonumber
\end{equation} 
\begin{equation}
\label{CF-sys1}
\sum_{k=1}^{I} \sigma_o^{-2}(y_k^o - x_k)^2  + \sigma_b^{-2}(x_0 - x_b)^2 
\end{equation} 
The control variable here is the entire trajectory within the assimilation interval $T$. In Eq. (\ref{CF-sys1}) we have used the fact that the model error bias, $\delta x^m(t)$, is given by $x(t) - x_0e^{\lambda t}$ assuming the model and the control trajectory, $x(t)$, are started from the same initial condition $x_0$. Note that $x_0$ is itself part of the estimation problem through the background term in the cost-function, and that the covariance matrices all reduce to scalar, such as $p_{t^{'}t^{''}} = p(t^{'},t^{''})$. 


While complete details can be found in \cite{CV10} we describe here the essential of the derivation.
The final minimizing solution of (\ref{CF-sys1}) is found using the technique of representer and reads:
\begin{equation}
\label{repr-sys1}
x(t)=x_be^{\lambda t} + \sum_{k=1}^{I}\beta_{k}r_{k}(t) = x^f(t) + \sum_{k=1}^{I}\beta_{k}r_{k}(t) \qquad 0\le t \le T
\end{equation}
The $I$ functions, $r_k(t)$, are the representers given by:
\begin{equation}
\label{repr-eq-sys1}
r_k(t) = r_k(0) e^{\lambda t} + \int_{0}^{T} p^{2}_{tt^{'}}(t^{'})a_k(t^{'})dt^{'}    \qquad 1\le k \le I 
\end{equation}
subject to $r_k(0)= \sigma_b^{2}\int_{0}^{T} a_k(t)e^{\lambda t} dt $, $1\le k \le I$, while the adjoint representers satisfy:
\begin{equation}
\label{reprADJ-eq-sys1}
a_k (t) = \delta(t-t_k)  \qquad 1\le k \le I 
\end{equation}
subject to $a_k(T)= 0$, $1\le k \le I$. The coefficients, $\beta_{k}$, are given by: 
\begin{equation}
\label{repr-coef}
{\bf \beta} = ({\bf S} + \sigma_o^{2}{\bf I}_d)^{-1}{\bf d}
\end{equation}
with ${\bf d}$ the innovation vector, ${\bf d}=(y_1^o-x^f_1,...,y_I^o-x^f_I)$, ${\bf S}$ the $I\times I$ matrix $({\bf S})_{i,j}=r_{i}(t_j)$, and ${\bf I}_d$ the $I\times I$ identity matrix. The coefficients are then inserted in (\ref{repr-sys1}) to obtain the final solution.

In the derivation of the general solution (\ref{repr-sys1}) (with the coefficients (\ref{repr-coef})), we have not specified the model error correlations $p^2(t^{'},t^{''})$; the particular choice adopted characterizes the formulations we aim to compare.
Our first choice consists in evaluating the model error correlations through (\ref{PMOD}).
By inserting $\delta\bm\mu=\frac{\partial{\bf f}}{\partial\lam}\delta\lam$, with $f(x)=\lambda x$, and the fundamental matrix, 
${\bf M}_{t,t_0}=e^{\lambda(t-t_0)}$, associated with the dynamics (\ref{sys1}), we get:
\begin{equation}
\label{mod_cov_sys1}
p^2(t^{'},t^{''}) = <(x_0\delta\lambda)^2> e^{\lambda (t^{'}+t^{''})}t^{'}t^{''}
\end{equation}
where the expectation, $< >$, is an average over a sample of initial conditions and parametric errors. Expression (\ref{mod_cov_sys1}) can now be inserted into (\ref{repr-eq-sys1})-(\ref{reprADJ-eq-sys1}), to obtain the $I$ representer functions in this case:
\begin{equation}
\label{repr-sol-sys1}
r_k(t)= e^{\lambda (t+t_{k})}[<(x_0\delta\lambda)^2>t_{k}t + \sigma_b^2] \qquad 1\le k \le M
\end{equation}
The representers (\ref{repr-sol-sys1}) are then inserted into (\ref{repr-coef}) to obtain the coefficients for the solution, $x(t)$, which is finally obtained through (\ref{repr-sys1}). This solution is hereafter referred to as the full weak-constraint. 

The same derivation is now repeated with the model error weights given by the short-time approximation (\ref{PMOD-app}). By substituting $\delta\bm\mu=\frac{\partial{\bf f}}{\partial\lam}\delta\lam$ into (\ref{PMOD-app}), we obtain: 
\begin{equation}
\label{PMOD-app-sys1}
p^2(t^{'},t^{''})= <(x_0\delta\lambda)^2>t^{'}t^{''}
\end{equation}
Once (\ref{PMOD-app-sys1}) is inserted into (\ref{repr-eq-sys1}) - (\ref{reprADJ-eq-sys1}) the representer solutions become: 
\begin{equation}
\label{repr-sol-sys1-app}
r_k(t)=\sigma_b^2e^{\lambda (t+t_{k})} + <(x_0\delta\lambda)^2>t_{k}t \qquad 1\le k \le I
\end{equation}
The representer functions are then introduced into (\ref{repr-coef}) and (\ref{repr-sys1}) to obtain the solution, $x(t)$, during the reference period $T$. The solution based on (\ref{repr-sol-sys1-app}) is the ST-w4DVar.

The strong-constraint solution is derived by invoking the continuity of the solution (\ref{repr-sol-sys1}), or (\ref{repr-sol-sys1-app}), with respect to the model error weights. The strong-constraint solution is obtained in the limit $\delta\lambda\rightarrow 0$, and reads:
\begin{equation}
\label{repr-sol-sys1-strong}
r_k(t)=\sigma_b^2e^{\lambda (t+t_{k})} \qquad 1\le k \le I
\end{equation}

\begin{figure}[h]
\begin{center}
\includegraphics[height=10cm,width=12.5cm]{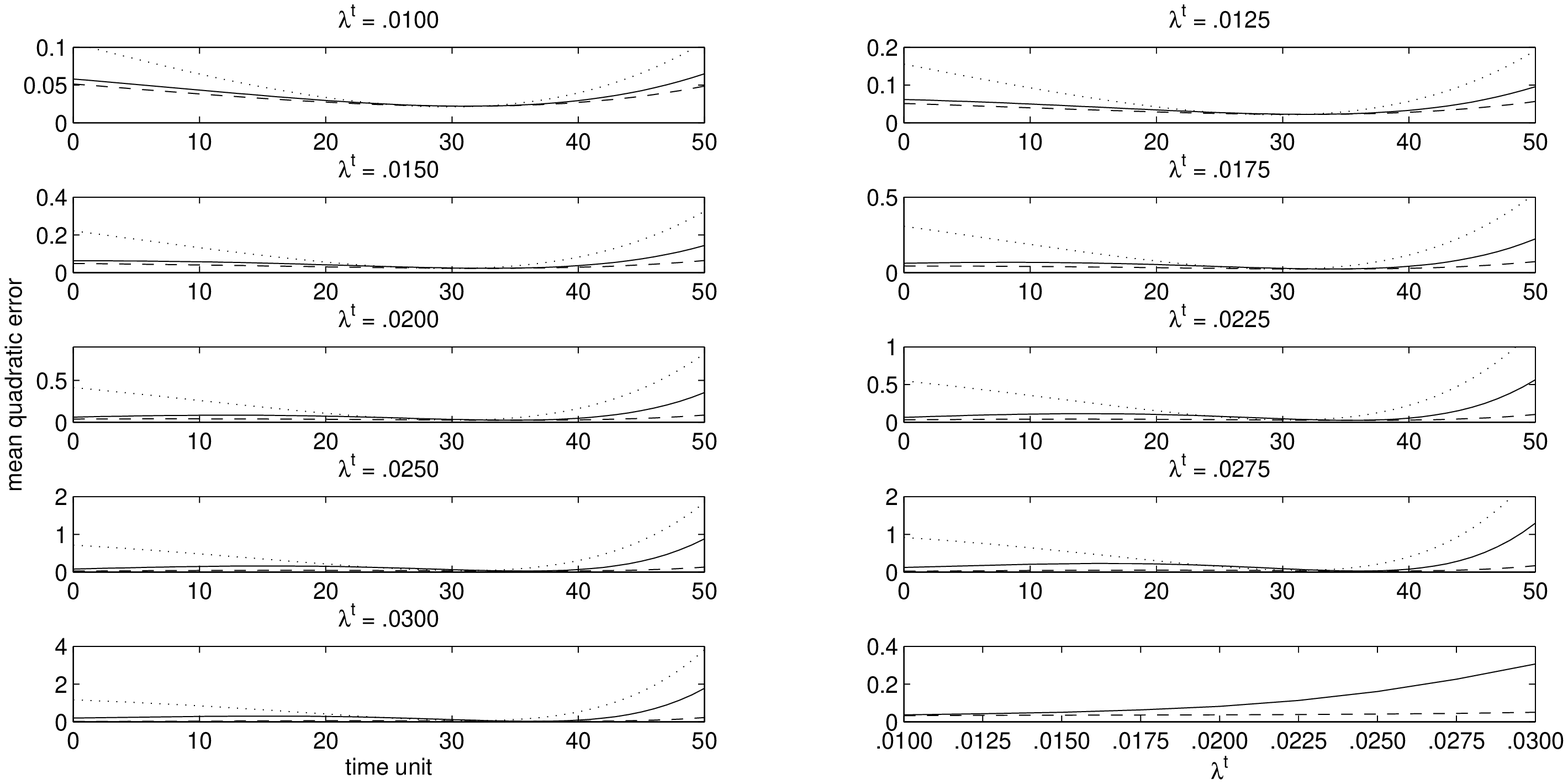}
\end{center}
\caption{Mean quadratic estimation error as a function of time for variational assimilation with system (\ref{sys1}). The panels refer to experiments with different $\lambda^{tr}$. The bottom-right panel shows the mean quadratic error, for the weak-constraint solutions only, averaged also over the assimilation interval $T$ as a function of $\lambda^{tr}$. Strong-constraint solution (dotted line), full weak-constraint solution (dashed line), short-time approximated weak-constraint solution (continuous line). From \cite{CV10}.}
\label{FIG7}
\end{figure}

The three solutions based respectively on (\ref{repr-sol-sys1}), (\ref{repr-sol-sys1-app}) and (\ref{repr-sol-sys1-strong}) are compared in Fig. \ref{FIG7}. Simulated noisy observations sampled from a Gaussian distribution around a solution of (\ref{sys1}), are distributed every $5$ time units over an assimilation interval $T=50$ time units. Different regimes of motion are considered by varying the true parameter $\lambda^{tr}$. The results displayed in Fig. \ref{FIG7} are averages over $10^3$ initial conditions and parametric model errors, around $x_0=2$ and $\lambda^{tr}$ respectively. The initial conditions are sampled from a Gaussian distribution with standard deviation $\sigma_b=1$, while the model parameter, $\lambda$, is sampled by a Gaussian distribution with standard deviation $\vert\Delta\lambda\vert = \vert \lambda^{tr} - \lambda \vert$; the observation error standard deviation is $\sigma_o=0.5$.

Figure \ref{FIG7} shows the mean quadratic estimation error, as a function of time, during the assimilation period $T$. The different panels refer to experiments with different parameter for the truth $0.01\le\lambda^{tr}\le0.03$, while the parametric error relative to the true value is set to $\Delta\lambda/\lambda^{tr}=50\%$. The three lines refer to the full weak-constraint (dashed line), ST-w4DVar (continuous line) and the strong-constraint (dotted line) solutions respectively. The bottom right panel summarizes the results and shows the mean error, averaged also in time, as a function of $\lambda^{tr}$ for the weak-constraint solutions only. As expected the full weak-constraint solution performs systematically better than any other approach. ST-w4DVar successfully outperforms the strong-constraint case, particularly at the beginning and end of the assimilation interval. The last plot displays the increase of total error of this solution as a function of $\lambda^{tr}$. To understand this dependence, one must recall that the duration of the short-time regime in a chaotic system is bounded by the inverse of the largest amplitude Lyapunov exponent \cite{Nicolis2003}. For the scalar unstable case considered here, this role is played by the parameter $\lambda^{tr}$. The increase of the total error of the short-time approximated weak-constraint as a function of $\lambda^{tr}$ reflects the progressive decrease of the accuracy of the short-time approximation for this fixed data assimilation interval, $T$.

\begin{figure}[h]
\begin{center}
\includegraphics[height=10cm,width=13cm]{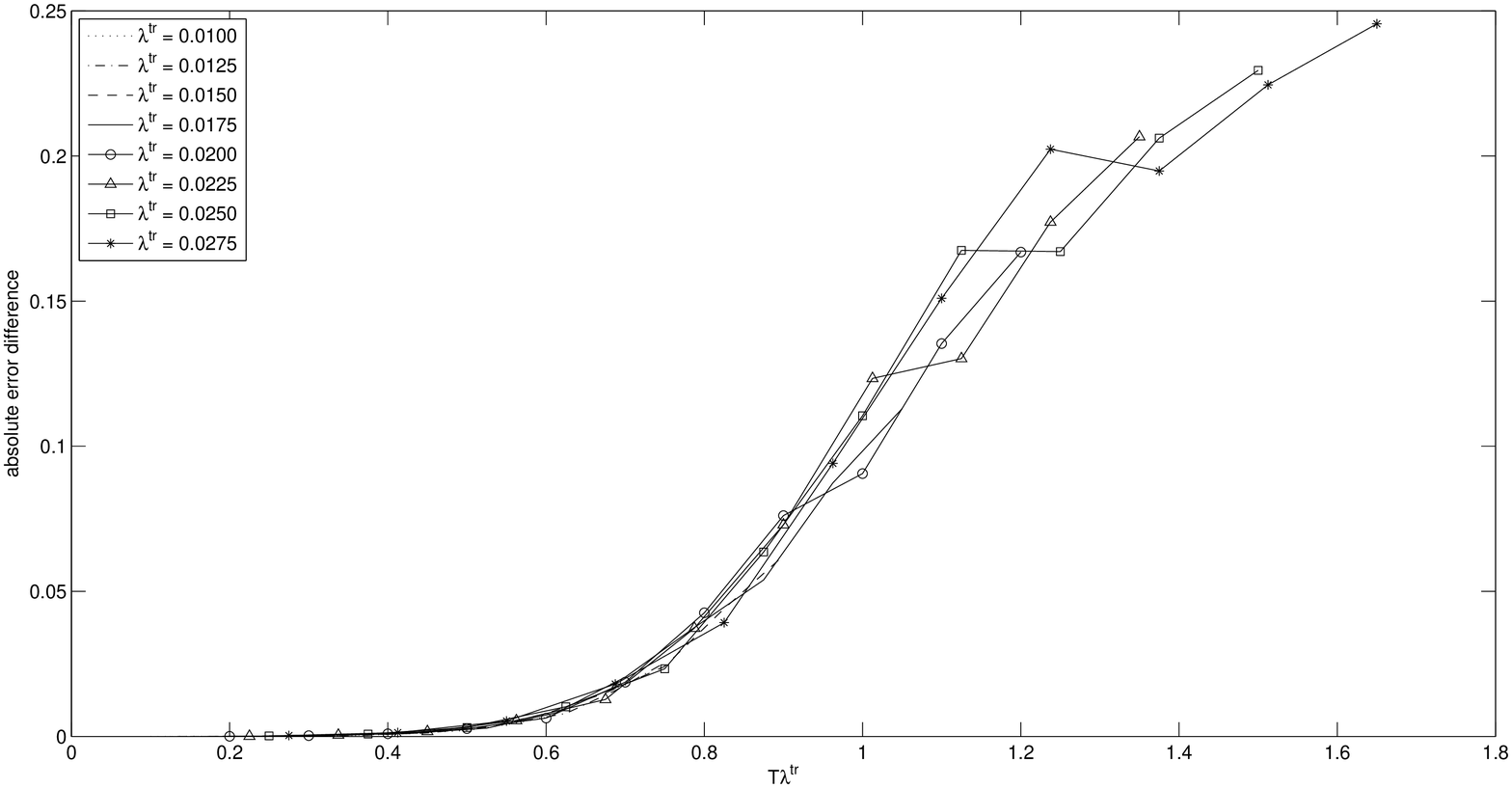}
\end{center}
\caption{Difference between the mean quadratic error of the short-time approximated and the full weak-constraint solution, for system (\ref{sys1}), as a function of $T\lambda^{tr}$, for assimilation period $10\le T\le 60$ and for different values of $\lambda^{tr}$. In all the experiments  $\Delta\lambda/\lambda^{tr}=50\%$. From \cite{CV10}.}
\label{FIG8}
\end{figure}

The accuracy of the ST-w4DVar in relation to the level of instability of the dynamics, is further summarized in Fig. \ref{FIG8}, where the difference between the mean quadratic error of this solution and the full weak-constraint one, is plotted as a function of the adimensional parameter $T\lambda^{tr}$, with $10\le T\le 60$ and $0.0100 \le \lambda^{tr} \le 0.0275 $. In all the experiments  $\Delta\lambda/\lambda^{tr}=50\%$. Remarkably all curves are superimposed, a clear indication that the accuracy of the analysis depends essentially on the product of the instability of the system and the data assimilation interval. 

We turn now to the case of a nonlinear dynamics. We adopt here the widely used Lorenz 3-variable convective system \cite{Lorenz1963}, whose equations read:
\begin{eqnarray}
\label{LOR}
\frac{dx}{dt}& =& -\sigma(x-y) \nonumber\\
\frac{dy}{dt}& =& \rho x -y-xz \\
\frac{dz}{dt}& =& xy-\beta z \nonumber
\end{eqnarray}
with $\lam=(\sigma, \rho, \beta)=(10,28,\frac{8}{3})$. OSSEs are performed with a solution of (\ref{LOR}) representing the reference dynamics from which observations are sampled. The estimation is based on observations of the entire system's state (${\it i.e.}$ observation operator equal to the identity $3\times 3$ matrix), distributed within a given assimilation interval and affected by an uncorrelated Gaussian error with covariance ${\bf R}$. The model dynamics is given by (\ref{LOR}) with a modified set of parameters. The numerical integrations are carried out with a second order Runge-Kutta scheme with a time-step equal to $0.01$ adimensional time units.

The variational cost function can be written, according to (\ref{CF}), as:
\begin{equation}
2J({\bf x}_0,{\bf x}_1,...,{\bf x}_L)= \sum_{i=1}^{L}\sum_{j=1}^{L}({\bf x}_i - \mathcal{M}({\bf x}_{i-1}))^T({\bf P}^m)^{-1}_{i,j}({\bf x}_j - \mathcal{M}({\bf x}_{j-1})) +  \nonumber
\end{equation}
\begin{equation}
\label{CF-2-Lorenz}
\sum_{k=1}^{I}({\bf y}_k^o-\mathcal{H}({\bf x}_k))^T{\bf R}^{-1}({\bf y}_k^o-\mathcal{H}({\bf x}_k)) + ({\bf x}_b - {\bf x}_0)^T{\bf B}^{-1}({\bf x}_b - {\bf x}_0) 
\end{equation}
We have assumed the assimilation interval $T$ has been discretized over $L$ time steps of equal length, $\Delta t$.

The control variable for the minimization is the series of the model state ${\bf x}_i$ at each time-step in the interval $T$. The minimizing solution is obtained by using a descent iterative method which makes use of the cost-function gradient with respect to ${\bf x}_i$, $0\le i \le L$. This latter reads: 
\begin{eqnarray}
\label{GRAD-2-Lorenz}
\mathbf{\nabla}_{{\bf x}_0}J&=&  -{\bf H}_0^T{\bf R}^{-1}({\bf y}_0^o -\mathcal{H}({\bf x}_0))  \nonumber \\
& & - {\bf M}_{0,1}^{T}[\sum_{j=1}^{L}({\bf P}^m_{1,j})^{-1}({\bf x}_j - \mathcal{M}({\bf x}_{j-1}))] - {\bf B}^{-1}({\bf x}_b - {\bf x}_0)  \qquad\qquad  i=0  \nonumber \\
\mathbf{\nabla}_{{\bf x}_i}J&=&  -{\bf H}_i^T{\bf R}^{-1}({\bf y}_i^o-\mathcal{H}({\bf x}_i)) - {\bf M}_{i,i+1}^{T}[\sum_{j=1}^{L}({\bf P}^m_{i+1,j})^{-1}({\bf x}_j - \mathcal{M}({\bf x}_{j-1}))] \nonumber \\
 & & + \sum_{j=1}^{L}{\bf P}_{i,j}^{-1}({\bf x}_j - \mathcal{M}({\bf x}_{j-1})) \qquad\qquad 1\le i \le L-1  \\
\mathbf{\nabla}_{{\bf x}_N}J&=&  -{\bf H}_N^T{\bf R}^{-1}({\bf y}_N^o-\mathcal{H}({\bf x}_N)) + \sum_{j=1}^{N}({\bf P}^m_{N,j})^{-1}({\bf x}_j - \mathcal{M}({\bf x}_{j-1})) \qquad\qquad i=L \nonumber
\end{eqnarray}

The gradient (\ref{GRAD-2-Lorenz}) is derived assuming that observations are available at each time step $t_i$, $0\le i \le L$. In the usual case of sparse observations the term proportional to the innovation disappears from the gradient with respect to the state vector at a time when observations are not present. 
Note furthermore that, if the model error is treated as an uncorrelated noise, the corresponding term in the cost-function reduces to a single summation over the time-steps weighted by the inverse of the model error covariances. The cost-function gradient modifies accordingly and the summation over all time-steps disappears \cite{Tremolet2006}.   

The cost-function (\ref{CF-2-Lorenz}) and its gradient (\ref{GRAD-2-Lorenz}) define the discrete weak-constraint variational problem. The ST-w4DVar consists in (\ref{CF-2-Lorenz}) and (\ref{GRAD-2-Lorenz}) with the model error correlations ${\bf P}^m_{i,j}$ estimated using the short-time approximation (\ref{PMOD-app}) that, in this discrete case, reads:
\begin{equation}
\label{PMOD-discr}
{\bf P}^m_{i,j} = <\delta\bm{\mu}_0\delta\bm{\mu}_0^T>ij \Delta t^2
\end{equation}
The invariant term $<\delta\bm\mu_0\delta\bm\mu_0^T>$, which is here a $3\times 3$ symmetric matrix, is assumed known a-priori and estimated by accumulating statistics on the model attractor, so that $<\delta\bm\mu_0\delta\bm\mu_0^T>=<<\delta\bm\mu_0\delta\bm\mu_0^T>>$ and perturbing randomly each of the three parameters $\sigma$, $\rho$ and $\beta$, with respect to the canonical values and with a standard deviation $\vert\Delta\lam\vert$. 
The ST-w4DVar is compared with the weak-contraint 4DVar with uncorrelated model error formulation and with the strong-constraint 4DVar; in this latter case the model error term disappears from the cost-function (\ref{CF-2-Lorenz}) and the gradient is computed with respect to the initial condition only \cite{TalagCourt1987}.   

\begin{figure}[h]
\begin{center}
\includegraphics[height=6.4cm,width=9.5cm]{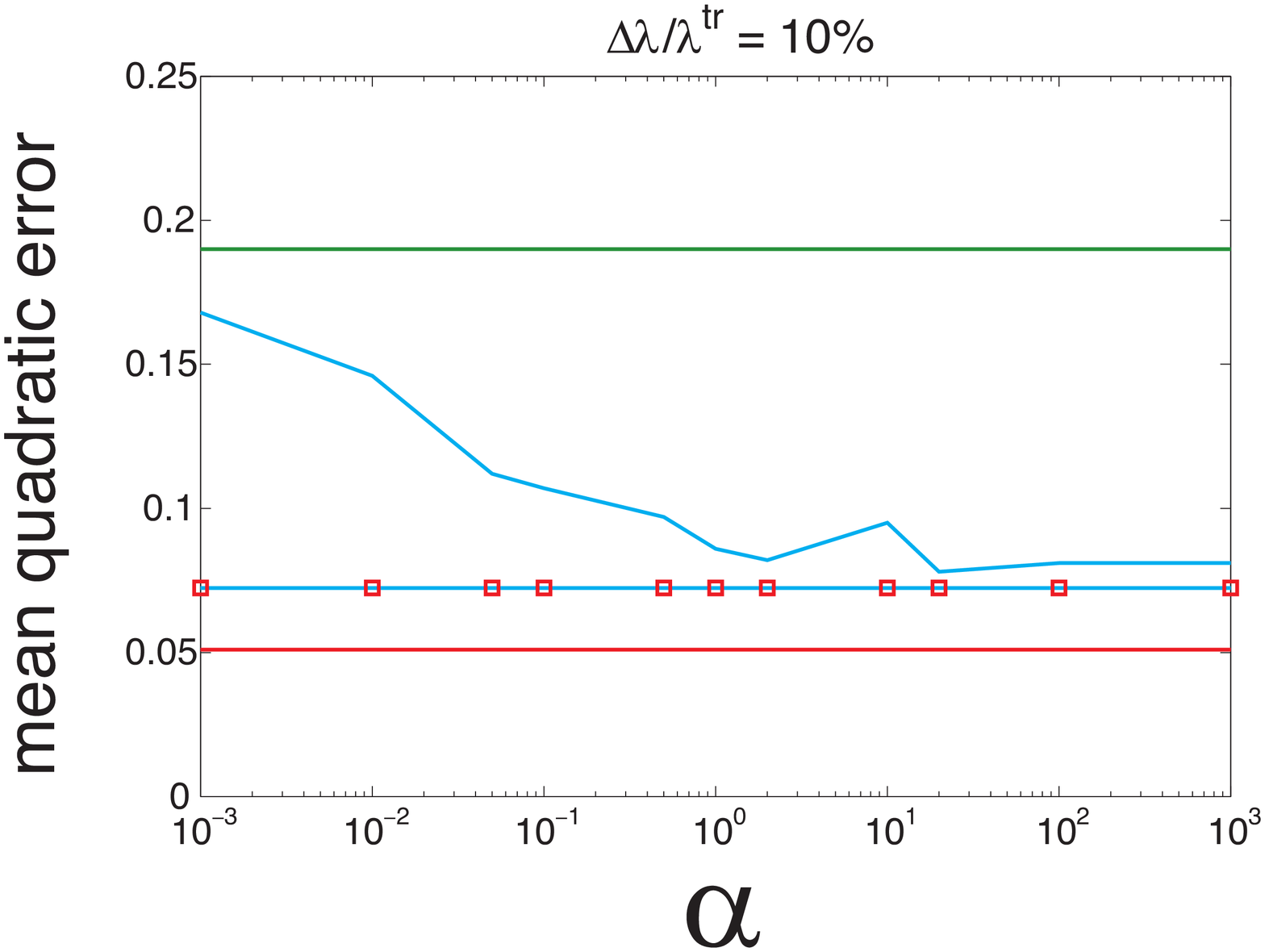}
\end{center}
\caption{Mean quadratic estimation error as a function of the tuning parameter $\alpha$ multiplying the model error covariance in the weak-constraint 4DVar with the uncorrelated noise assumption (see text for details). The dynamics is given by system (\ref{LOR}). ST-w4DVar (red), strong-constraint 4DVar (green), uncorrelated noise weak-constraint 4DVar (blue) and uncorrelated noise weak-constraint 4DVar with spatial covariance as in the short-time approximated weak-constraint (blue with red marks). From \cite{CV10}.}
\label{FIG9}
\end{figure}

The assumption of uncorrelated model error is done often in applications. It is particularly attractive in view of the consequent reduction of the computational cost associated with the minimization procedure. Model error covariance are commonly modeled as proportional to the background matrix, so that ${\bf P}^m = \alpha{\bf B}$.  
Figure \ref{FIG9} shows the mean quadratic error as a function of the tuning parameter. Results are averaged over an ensemble of $50$ initial conditions and parametric model error; the observation and assimilation interval are set to $2$ and $8$ time-steps respectively. Strong constraint 4DVar (green) and ST-w4DVar (red) do not depend on $\alpha$ and are therefore horizontal lines in the panel. Weak constraint 4DVar with uncorrelated model error (blue) shows, as expected, a marked dependence on the model error covariance amplitude.   
The blue line with squared marks refers to an experiment where the model error is treated as an uncorrelated noise but the spatial covariances at observing times are estimated using the short-time approximation. Comparing this curve with the blue line relative to weak constraint 4DVar with uncorrelated model error and ${\bf P}^m = \alpha{\bf B}$ allow for evaluating the impact of neglecting the time correlation and of using an incorrect spatial covariance.

The uncorrelated noise formulation (solid line with no marks) never reaches the accuracy of the ST-w4DVar. Note furthermore that for small $\alpha$ it almost reaches the same error level as the strong-constraint 4DVar where the model is assumed to be perfect. By further increasing $\alpha$ over $\alpha=10^3$ (not shown) the error reaches a plateau whose value is controlled by the observation error level. When the spatial covariances are estimated as in the short-time weak-constraint the performance is generally improved, although for large $\alpha$, the estimate ${\bf P}^m=\alpha{\bf B}$ gives very close skill and the improvement in correspondence with the best-possible $\alpha$ is only minor. This suggests that the degradation of the uncorrelated noise formulation over the short-time weak-constraint is mainly the consequence of neglecting the time correlation and only to a small extent to the use of an incorrect spatial covariance.

\section{Discussion\label{sec:discussion}}

Data assimilation schemes are usually assuming the uncorrelated nature of model uncertainties. This choice is indeed legitimate as a first
approximation when initial condition errors dominate model errors. Due to the large increase of measurement data availability and quality, 
this view should be reassessed. However one prominent difficulty in dealing with model errors is the wide variety of potential sources of
uncertainties, going from parametric errors up to the absence of description of some dynamical processes. But recently a stream of works, \cite{Nicolis2003,Nicolis2004,Nicolis2009},
provided important insights into the dynamics of deterministic model uncertainties, and from which generic mechanisms of growth 
were disentangled. In particular, it was shown that the mean square error associated with the presence of deterministic model errors is 
growing quadratically in time at short lead times. These insights now open new avenues in the description of data assimilation schemes, as
it has been demonstrated in the present chapter.    

First, the deterministic approach was applied in the context of the Extended Kalman Filter, for both the classical state estimation
scheme and its augmented version (state and parameters). It has been demonstrated that these new schemes are indeed most valuable when dealing
with deterministic model error sources. They provide a large improvement in the state estimation (in particular with the ST-AEKF) not only
in the context of idealized settings (Lorenz' system) but as also for realistic applications as shown by the results obtained
with the offline version of an operational soil model (ISBA). 
Second the same idea was investigated in the context of four dimensional variational assimilation for which a weak-constrained framework 
was adopted. In 
this case model error cross-correlations were considered as quantities depending quadratically on time, implying a time dependent weighting
of the model error terms during the assimilation period. This approach also led to important improvements as compared to more drastic
assumptions like the time independence of model error source terms, or the absence of such sources.     

The approaches proposed in this chapter rely on an important assumption, the deterministic nature of model uncertainties. As alluded in Section 2, the proposed general setting could also be extended to independent random noises, provided the appropriate temporal variations of the
bias (Eq. 5) and covariances (Eq. 7) are used (see the remark at the end of Section 2). Still the covariance will be dependent on time (linearly) (see {\it e.g.} \cite{Vannitsem2002}) 
and will for instance affect differently the weights of the weak constrained four dimensional variational assimilation. Besides these 
technical aspects,
it remains difficult to know what the exact nature of the sources of model errors is. A realistic view would be that the fast time scale
processes -- like turbulence in the surface boundary layer -- could be considered as random components, while parameterization errors in 
the larger 
scale description of the stability of the atmosphere is a deterministic process. This leads us to consider model errors at a certain scale of 
description as a deterministic component plus a random component, and the user is left with the delicate question of evaluating whether the
dominant sources present in his problem are of one kind or the other.         

This work has mostly tackled this problem in the context of simple idealized dynamical systems. These encouraging results
need however to be confronted with more operational problems. In line with the results obtained with soil model, this problem is currently under 
investigation in the context of the use of the ST-AEKF for an online version of ISBA coupled with ALARO at the Royal Meteorological 
Institute of Belgium. 
A relevant methodological development will be the incorporation of the deterministic model error treatment in the context of the ensemble based schemes, as illustrated in \cite{MitchellCarrassi2015} or \cite{Raanes_et_al_2015}. Finally, it is worth mentioning that the deterministic model error dynamics has been recently used in a new drift correction procedure in the context of interannual-to-decadal predictions \cite{CWGDAV2014}. 
These recent applications illustrate the usefulness of the deterministic approach and should be further extended to a wider range of applications in which model error is present. In particular in the context of coupled atmosphere-ocean systems where multiple scales of motion are present and model error often originates at the level of the coupling.

\end{document}